\documentclass[preprint,superscriptaddress,nofootinbib]{revtex4}
  
  \usepackage{graphicx}
  \begin{document}
  \title{${\Upsilon}(nS)$ ${\to}$ $B_{c}^{\ast}D$ decays with perturbative QCD approach}
  \author{Yueling Yang}
  \affiliation{Institute of Particle and Nuclear Physics,
              Henan Normal University, Xinxiang 453007, China}
  \author{Junfeng Sun}
  \affiliation{Institute of Particle and Nuclear Physics,
              Henan Normal University, Xinxiang 453007, China}
  \author{Jinshu Huang}
  \affiliation{College of Physics and Electronic Engineering,
              Nanyang Normal University, Nanyang 473061, China}
  \author{Haiyan Li}
  \affiliation{Institute of Particle and Nuclear Physics,
              Henan Normal University, Xinxiang 453007, China}
  \author{Gongru Lu}
  \affiliation{Institute of Particle and Nuclear Physics,
              Henan Normal University, Xinxiang 453007, China}
  \author{Qin Chang}
  \affiliation{Institute of Particle and Nuclear Physics,
              Henan Normal University, Xinxiang 453007, China}
  %%%%%%%%%%%%%%%%%%%%%%%%%%%%%%
  \begin{abstract}
  The ${\Upsilon}(nS)$ ${\to}$ $B_{c}^{\ast}D$ weak decays
  ($n$ $=$ $1$, $2$, $3$) are investigated with perturbative QCD approach.
  It is found that the CKM-favored ${\Upsilon}(nS)$ ${\to}$
  $B_{c}^{\ast}D_{s}$ decays have branching ratio of ${\cal O}(10^{-10})$,
  which might be potentially accessible to the future LHC and
  SuperKEKB experiments.
  \end{abstract}
  \maketitle

  %%%%%%%%%%%%%%%%%%%%%%%%%%%%%%
  \section{Introduction}
  \label{sec01}
  ${\Upsilon}(1S)$, ${\Upsilon}(2S)$ and ${\Upsilon}(3S)$ are
  spin-triplet $S$-wave $b\bar{b}$ bound states carrying with
  quantum number of $I^{G}J^{PC}$ $=$ $0^{-}1^{--}$ \cite{pdg}.
  They all lie below the open bottom threshold. They must
  strongly decay into two light hadrons via $b\bar{b}$
  annihilation into at least three gluons.
  So their decay width is very narrow, only dozens of keV.
  [Hereinafter, for simplicity sake, we will use a notation
  ${\Upsilon}(nS)$ to represent ${\Upsilon}(1S)$, ${\Upsilon}(2S)$,
  and ${\Upsilon}(3S)$ mesons.]
  Since their discovery in 1977 \cite{herb,innes}, ${\Upsilon}(nS)$
  has been attracting much attention from experimentalists
  and theorists.
  Thanks to the excellent performance from experimental groups of CLEO,
  BaBar, Belle, CDF, D0, LHCb, ATLAS and so on, remarkable achievements
  have been made in understanding of the nature of upsilon \cite{pdg,1212.6552}.
  The strong and electromagnetic ${\Upsilon}(nS)$ decay modes have
  been carefully investigated.
  With accumulation of ${\Upsilon}(nS)$ data samples,
  it might be possible to search for ${\Upsilon}(nS)$
  weak decay at future LHC and SuperKEKB experiments.

  Theoretically, both valence quarks of ${\Upsilon}(nS)$ can
  decay individually via the weak interaction.
  The $b$ ${\to}$ $c$ transition is particularly favored by
  a hierarchy of the Cabibbo-Kabayashi-Maskawa (CKM) matrix
  elements. So ${\Upsilon}(nS)$ decay into final states
  containing a $B_{c}^{(\ast)}$ meson should, in principle,
  have a relatively large branching fraction among its weak
  decay modes.
  Recently, some phenomenological QCD-inspired methods have
  been vividly developed to deal with heavy quark weak decay,
  such as perturbative QCD (pQCD) approach \cite{pqcd1,pqcd2,pqcd3},
  QCD factorization (QCDF) \cite{qcdf1,qcdf2,qcdf3,plb488,prd64}
  and soft and collinear effective theory
  \cite{scet1,scet2,scet3,scet4}. The
  ${\Upsilon}(nS)$ ${\to}$ $B_{c}^{\ast}D$ decays offer
  a good plaza to ulteriorly test various phenomenological
  models and to further explore the underlying dynamical
  mechanism of heavy quarkonium weak decay.
  In addition, as far as we know,
  there is no experimental measurement report and few
  theoretical work related to ${\Upsilon}(nS)$ ${\to}$
  $B_{c}^{\ast}D$ decay for the moment.
  Herein, we will study the bottom- and charm-changing
  ${\Upsilon}(nS)$ ${\to}$ $B_{c}^{\ast}D$ weak decays with
  pQCD approach to provide future experimental exploration
  with a useful reference.

  This paper is organized as follows.
  The section \ref{sec02} devotes to theoretical framework
  and amplitudes for ${\Upsilon}(nS)$ ${\to}$ $B_{c}^{\ast}D$
  decay. The numerical results and discussion are presented
  in section \ref{sec03}.
  We summarize in the last section.

  %%%%%%%%%%%%%%%%%%%%%%%%%%%%%%
  \section{theoretical framework}
  \label{sec02}
  %%%%%%%%%%%%%%%%%%%%%%%%%%
  \subsection{The effective Hamiltonian}
  \label{sec0201}
  The effective weak Hamiltonian describing nonleptonic
  ${\Upsilon}(nS)$ ${\to}$ $B_{c}^{\ast}D$ decays is
  written as \cite{9512380}
 %---------------------------------------------------------
   \begin{equation}
  {\cal H}_{\rm eff}\ =\
   \frac{G_{F}}{\sqrt{2}}\, \Big\{
   V_{cb} V_{cq}^{\ast}
   \sum\limits_{i=1}^{2} C_{i}({\mu})\,Q_{i}({\mu})
  -V_{tb} V_{tq}^{\ast}
   \sum\limits_{j=3}^{10}
   C_{j}({\mu})\,Q_{j}({\mu}) \Big\}
   + {\rm h.c.}
   \label{hamilton},
   \end{equation}
 %---------------------------------------------------------
  where $G_{F}$ ${\simeq}$ $1.166{\times}10^{-5}\,{\rm GeV}^{-2}$
  is the Fermi constant \cite{pdg}; $q$ $=$ $d$ and $s$;
  the CKM factors can be expressed as
 %---------------------------------------------------------
  \begin{eqnarray}
  V_{cb}V_{cs}^{\ast} &=&
  +            A{\lambda}^{2}
  - \frac{1}{2}A{\lambda}^{4}
  - \frac{1}{8}A{\lambda}^{6}(1+4A^{2})
  +{\cal O}({\lambda}^{7})
  \label{vcbcs}, \\
  V_{tb}V_{ts}^{\ast} &=& -V_{cb}V_{cs}^{\ast}
  - A{\lambda}^{4}({\rho}-i{\eta})
  +{\cal O}({\lambda}^{7})
  \label{vtbts},
  \end{eqnarray}
 %---------------------------------------------------------
  for ${\Upsilon}(nS)$ ${\to}$ $B_{c}^{{\ast}}D_{s}$
  decays; and
 %---------------------------------------------------------
  \begin{eqnarray}
  V_{cb}V_{cd}^{\ast} &=& -A{\lambda}^{3}
  +{\cal O}({\lambda}^{7})
  \label{vcbcd}, \\
  V_{tb}V_{td}^{\ast} &=& +A{\lambda}^{3}(1-{\rho}+i{\eta})
  + \frac{1}{2}A{\lambda}^{5}({\rho}-i{\eta})
  +{\cal O}({\lambda}^{7})
  \label{vtbtd},
  \end{eqnarray}
 %---------------------------------------------------------
  for ${\Upsilon}(nS)$ ${\to}$ $B_{c}^{({\ast})}D_{d}$
  decays; $A$, ${\lambda}$, ${\rho}$ and ${\eta}$ are
  Wolfenstein parameters \cite{pdg,wolfenstein}.

  The local tree operators $Q_{1,2}$,
  QCD penguin operators $Q_{3,{\cdots},6}$,
  and electroweak operators $Q_{7,{\cdots},10}$ are defined below.
 %-----------------------------------------------------
    \begin{eqnarray}
    Q_{1} &=&
  [ \bar{c}_{\alpha}{\gamma}_{\mu}(1-{\gamma}_{5})b_{\alpha} ]
  [ \bar{q}_{\beta} {\gamma}^{\mu}(1-{\gamma}_{5})c_{\beta} ]
    \label{q1}, \\
 %-----------------------------------------------------
    Q_{2} &=&
  [ \bar{c}_{\alpha}{\gamma}_{\mu}(1-{\gamma}_{5})b_{\beta} ]
  [ \bar{q}_{\beta}{\gamma}^{\mu}(1-{\gamma}_{5})c_{\alpha} ]
    \label{q2},
    \end{eqnarray}
 %-----------------------------------------------------
    \begin{eqnarray}
    Q_{3} &=& \sum\limits_{q^{\prime}}
  [ \bar{q}_{\alpha}{\gamma}_{\mu}(1-{\gamma}_{5})b_{\alpha} ]
  [ \bar{q}^{\prime}_{\beta} {\gamma}^{\mu}(1-{\gamma}_{5})q^{\prime}_{\beta} ]
    \label{q3}, \\
 %-----------------------------------------------------
    Q_{4} &=& \sum\limits_{q^{\prime}}
  [ \bar{q}_{\alpha}{\gamma}_{\mu}(1-{\gamma}_{5})b_{\beta} ]
  [ \bar{q}^{\prime}_{\beta}{\gamma}^{\mu}(1-{\gamma}_{5})q^{\prime}_{\alpha} ]
    \label{q4}, \\
 %-----------------------------------------------------
    Q_{5} &=& \sum\limits_{q^{\prime}}
  [ \bar{q}_{\alpha}{\gamma}_{\mu}(1-{\gamma}_{5})b_{\alpha} ]
  [ \bar{q}^{\prime}_{\beta} {\gamma}^{\mu}(1+{\gamma}_{5})q^{\prime}_{\beta} ]
    \label{q5}, \\
 %-----------------------------------------------------
    Q_{6} &=& \sum\limits_{q^{\prime}}
  [ \bar{q}_{\alpha}{\gamma}_{\mu}(1-{\gamma}_{5})b_{\beta} ]
  [ \bar{q}^{\prime}_{\beta}{\gamma}^{\mu}(1+{\gamma}_{5})q^{\prime}_{\alpha} ]
    \label{q6},
    \end{eqnarray}
 %-----------------------------------------------------
    \begin{eqnarray}
    Q_{7} &=& \sum\limits_{q^{\prime}} \frac{3}{2}e_{q^{\prime}}\,
  [ \bar{q}_{\alpha}{\gamma}_{\mu}(1-{\gamma}_{5})b_{\alpha} ]
  [ \bar{q}^{\prime}_{\beta} {\gamma}^{\mu}(1+{\gamma}_{5})q^{\prime}_{\beta} ]
    \label{q7}, \\
 %-----------------------------------------------------
    Q_{8} &=& \sum\limits_{q^{\prime}} \frac{3}{2}e_{q^{\prime}}\,
  [ \bar{q}_{\alpha}{\gamma}_{\mu}(1-{\gamma}_{5})b_{\beta} ]
  [ \bar{q}^{\prime}_{\beta}{\gamma}^{\mu}(1+{\gamma}_{5})q^{\prime}_{\alpha} ]
    \label{q8}, \\
 %-----------------------------------------------------
    Q_{9} &=& \sum\limits_{q^{\prime}} \frac{3}{2}e_{q^{\prime}}\,
  [ \bar{q}_{\alpha}{\gamma}_{\mu}(1-{\gamma}_{5})b_{\alpha} ]
  [ \bar{q}^{\prime}_{\beta} {\gamma}^{\mu}(1-{\gamma}_{5})q^{\prime}_{\beta} ]
    \label{q9}, \\
 %-----------------------------------------------------
    Q_{10} &=& \sum\limits_{q^{\prime}} \frac{3}{2}e_{q^{\prime}}\,
  [ \bar{q}_{\alpha}{\gamma}_{\mu}(1-{\gamma}_{5})b_{\beta} ]
  [ \bar{q}^{\prime}_{\beta}{\gamma}^{\mu}(1-{\gamma}_{5})q^{\prime}_{\alpha} ]
    \label{q10},
    \end{eqnarray}
 %-----------------------------------------------------
  where ${\alpha}$ and ${\beta}$ are color indices;
  $q^{\prime}$ $=$ $u$, $d$, $s$, $c$, $b$ has an
  electric charge $e_{q^{\prime}}$ in the unit
  of ${\vert}e{\vert}$.

  The scale ${\mu}$ separates physical contributions into two components.
  The Wilson coefficients $C_{i}(\mu)$ summarize the physical contributions
  above ${\mu}$, and has been reliably computed to the next-to-leading
  order with perturbation theory \cite{9512380}.
  The hadronic matrix elements (HME), where the local operators are
  sandwiched between initial and final hadron states, contain the physical
  contributions below ${\mu}$.
  Due to the incorporation of long distance contributions and the
  entanglement of perturbative and nonperturbative effects,
  HME is not yet fully understood until now.
  However, in order to evaluate the amplitudes, one has to face
  directly the HME's calculation based on some approximation and
  assumptions, which leads to large theoretical uncertainties.

  %%%%%%%%%%%%%%%%%%%%%%%%%%
  \subsection{Hadronic matrix elements}
  \label{sec0202}
  Phenomenologically, combining factorization hypothesis
  \cite{npb133,zpc29,npbps11} and hard-scattering approach
  \cite{plb87,prd22,plb90,prd21,plb94}, HME could be written
  as the convolution of hard scattering kernel function ${\cal T}$
  and distribution amplitudes (DAs) of participating hadrons.
  DAs are nonperturbative but universal inputs, which can be
  obtained from nonperturbative methods or fitted from
  experimental data.
  In order to eliminate the endpoint singularities accompanying
  with spectator rescattering and annihilation contributions
  based on a collinear approximation \cite{qcdf3,plb488,prd64},
  and in the meantime to provide an effective cutoff on
  nonperturbative contributions, the transverse momentum of
  valence quarks is kept explicitly and a Sudakov factor for
  each of DAs is introduced compulsorily with pQCD approach
  \cite{pqcd1,pqcd2,pqcd3}.
  A general pQCD amplitude is made up of three parts:
  the Wilson coefficients $C_{i}$ absorbing physical contributions
  above a typical scale of $t$, hard scattering kernel
  function ${\cal T}$ accounting for heavy quark weak decay,
  and wave functions ${\Phi}$, i.e.,
  %-----------------------------------------------------
  \begin{equation}
  {\int} dk\,
  C_{i}(t)\,{\cal T}(t,k)\,\prod_{j}{\Phi}_{j}(k)\,e^{-S_{j}}
  \label{hadronic},
  \end{equation}
  %-----------------------------------------------------
  where $k$ is the momentum of valence quarks, and
  $e^{-S_{j}}$ is a Sudakov factor.

  %%%%%%%%%%%%%%%%%%%%%%%%%%
  \subsection{Kinematic variables}
  \label{sec0203}
\begin{figure}[h]
  \includegraphics[width=0.98\textwidth,bb=80 510 530 710]{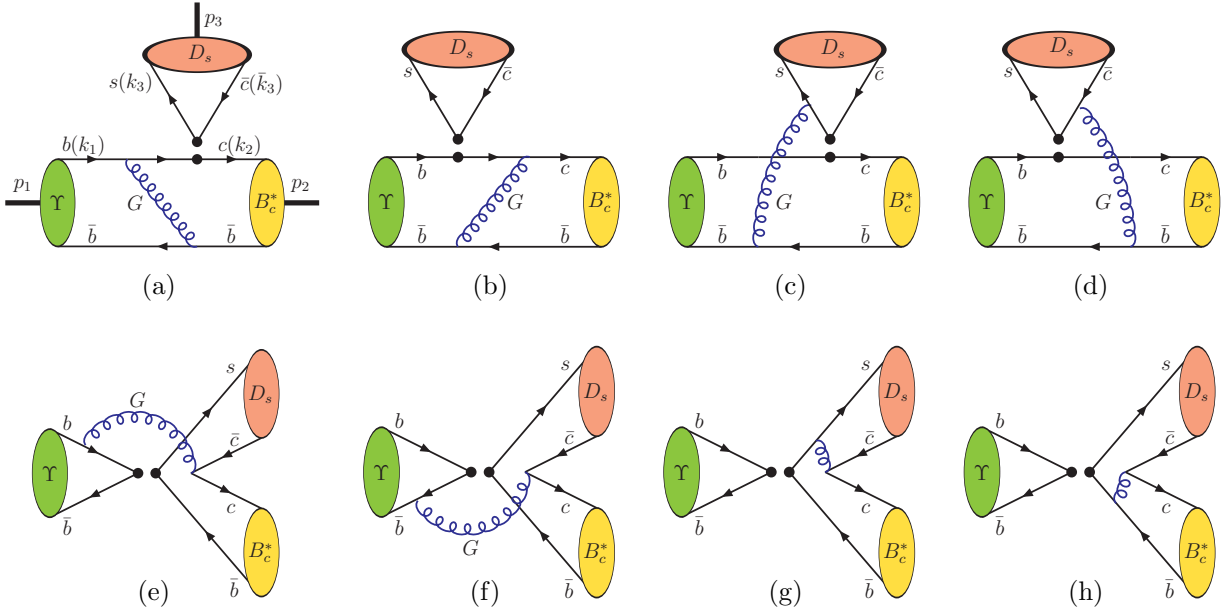}
  \caption{Feynman diagrams for ${\Upsilon}(nS)$ ${\to}$
  $B_{c}^{\ast}D_{s}$ decay with pQCD approach, where
  (a,b) are factorizable emission diagrams,
  (c,d) are nonfactorizable emission diagrams,
  (e,f) are nonfactorizable annihilation diagrams,
  and (g,h) are factorizable annihilation diagrams.}
  \label{fig:fey}
  \end{figure}

  In the ${\Upsilon}(nS)$ rest frame,
  the light cone kinematic variables are defined as follows.
  %------------------------------------
  \begin{equation}
  p_{\Upsilon}\, =\, p_{1}\, =\, \frac{m_{1}}{\sqrt{2}}(1,1,0)
  \label{kine-p1},
  \end{equation}
  %------------------------------------
  \begin{equation}
  p_{B_{c}^{\ast}}\, =\, p_{2}\, =\, (p_{2}^{+},p_{2}^{-},0)
  \label{kine-p2},
  \end{equation}
  %------------------------------------
  \begin{equation}
  p_{D}\, =\, p_{3}\, =\, (p_{3}^{-},p_{3}^{+},0)
  \label{kine-p3},
  \end{equation}
  %------------------------------------
  \begin{equation}
  p_{i}^{\pm}\, =\, (E_{i}\,{\pm}\,p)/\sqrt{2}
  \label{kine-pipm},
  \end{equation}
  %------------------------------------
  \begin{equation}
  k_{i}\, =\, x_{i}\,p_{i}+(0,0,\vec{k}_{iT})
  \label{kine-ki},
  \end{equation}
  %------------------------------------
  \begin{equation}
 {\epsilon}_{1}^{\parallel}\, =\,
  \frac{p_{1}}{m_{1}}-\frac{m_{1}}{p_{1}{\cdot}n_{+}}n_{+}
  \label{kine13-longe},
  \end{equation}
  %------------------------------------
  \begin{equation}
 {\epsilon}_{2}^{\parallel}\, =\,
  \frac{p_{2}}{m_{2}}-\frac{m_{2}}{p_{2}{\cdot}n_{-}}n_{-}
  \label{kine-longe},
  \end{equation}
  %------------------------------------
  \begin{equation}
 {\epsilon}_{1,2}^{\perp}\, =\, (0,0,\vec{1})
  \label{kine-transe},
  \end{equation}
  %------------------------------------
  \begin{equation}
  n_{+}=(1,0,0)
  \label{kine-nullp},
  \end{equation}
  %------------------------------------
  \begin{equation}
  n_{-}=(0,1,0)
  \label{kine-nullm},
  \end{equation}
  %------------------------------------
  \begin{equation}
  s\, =\, 2\,p_{2}{\cdot}p_{3}\, =\, m_{1}^{2}-m_{2}^{2}-m_{3}^{2}
  \label{kine-s},
  \end{equation}
  %------------------------------------
  \begin{equation}
  t\, =\, 2\,p_{1}{\cdot}p_{2}\, =\, m_{1}^{2}+m_{2}^{2}-m_{3}^{2}\, =\, 2\,m_{1}\,E_{2}
  \label{kine-t},
  \end{equation}
  %------------------------------------
  \begin{equation}
  u\, =\, 2\,p_{1}{\cdot}p_{3}\, =\, m_{1}^{2}-m_{2}^{2}+m_{3}^{2}\, =\, 2\,m_{1}\,E_{3}
  \label{kine-u},
  \end{equation}
  %------------------------------------
  \begin{equation}
  u+t-s\, =\, m_{1}^{2}+m_{2}^{2}+m_{3}^{2}
  \label{kine-s-t-u},
  \end{equation}
  %------------------------------------
  \begin{equation}
  s\,t+s\,u-t\,u-4\,m_{1}^{2}\,p^{2}\, =\, 0
  \label{kine-pcm},
  \end{equation}
  %------------------------------------
  where $x_{i}$ and $k_{iT}$ are the longitudinal momentum
  fraction and transverse momentum of valence quark, respectively;
  ${\epsilon}_{i}^{\parallel}$ and ${\epsilon}_{i}^{\perp}$ are the
  longitudinal and transverse polarization vectors, respectively,
  satisfying relations ${\epsilon}_{i}^{2}$ $=$ $-1$ and
  ${\epsilon}_{i}{\cdot}p_{i}$ $=$ $0$;
  the subscript $i$ $=$ $1$, $2$, $3$ on variables ($E_{i}$, $p_{i}$,
  $m_{i}$, ${\epsilon}_{i}$) corresponds to
  ${\Upsilon}(nS)$, $B_{c}^{\ast}$, and $D$ mesons, respectively;
  $n_{+}$ and $n_{-}$ are the positive and negative null vectors,
  respectively; $s$, $t$ and $u$ are Lorentz-invariant variables.
  These kinematic variables are showed in Fig.\ref{fig:fey}(a).

  %%%%%%%%%%%%%%%%%%%%%%%%%%
  \subsection{Wave functions}
  \label{sec0204}
  The definitions of wave functions are \cite{prd65,jhep0703},
  %------------------------------------
  \begin{equation}
 {\langle}0{\vert}b_{i}(z)\bar{b}_{j}(0){\vert}
 {\Upsilon}(p_{1},{\epsilon}_{1}^{{\parallel}}){\rangle}\,
 =\, \frac{f_{{\Upsilon}}}{4}{\int}dk_{1}\,e^{-ik_{1}{\cdot}z}
  \Big\{ \!\!\not{\epsilon}_{1}^{{\parallel}} \Big[
   m_{1}\,{\phi}_{\Upsilon}^{v}(k_{1})
  -\!\!\not{p}_{1}\, {\phi}_{\Upsilon}^{t}(k_{1})
  \Big] \Big\}_{ji}
  \label{wave-bb-long},
  \end{equation}
  %------------------------------------
  \begin{equation}
 {\langle}0{\vert}b_{i}(z)\bar{b}_{j}(0){\vert}
 {\Upsilon}(p_{1},{\epsilon}_{1}^{{\perp}}){\rangle}\,
 =\, \frac{f_{{\Upsilon}}}{4}{\int}dk_{1}\,e^{-ik_{1}{\cdot}z}
  \Big\{ \!\!\not{\epsilon}_{1}^{{\perp}} \Big[
   m_{1}\,{\phi}_{\Upsilon}^{V}(k_{1})
  -\!\!\not{p}_{1}\, {\phi}_{\Upsilon}^{T}(k_{1})
  \Big] \Big\}_{ji}
  \label{wave-bb-perp},
  \end{equation}
  %------------------------------------
  %------------------------------------
  \begin{equation}
 {\langle}B_{c}^{\ast}(p_{2},{\epsilon}_{2}^{{\parallel}})
 {\vert}\bar{c}_{i}(z)b_{j}(0){\vert}0{\rangle}\ =\
  \frac{f_{B_{c}^{\ast}}}{4}{\int}_{0}^{1}dk_{3}\,e^{ik_{2}{\cdot}z}
  \Big\{ \!\not{\epsilon}_{2}^{{\parallel}} \Big[
   m_{2}\,{\phi}_{B_{c}^{\ast}}^{v}(k_{2})
  +\!\not{p}_{2}\, {\phi}_{B_{c}^{\ast}}^{t}(k_{2})
  \Big] \Big\}_{ji}
  \label{wave-bc-long},
  \end{equation}
  %------------------------------------
  \begin{equation}
 {\langle}B_{c}^{{\ast}}(p_{2},{\epsilon}_{2}^{{\perp}})
 {\vert}\bar{c}_{i}(z)b_{j}(0){\vert}0{\rangle}\ =\
  \frac{f_{B_{c}^{\ast}}}{4}{\int}_{0}^{1}dk_{2}\,e^{ik_{2}{\cdot}z}
  \Big\{ \!\not{\epsilon}_{2}^{{\perp}} \Big[
   m_{2}\,{\phi}_{B_{c}^{\ast}}^{V}(k_{2})
 +\!\not{p}_{2}\, {\phi}_{B_{c}^{\ast}}^{T}(k_{2})
  \Big] \Big\}_{ji}
  \label{wave-bc-perp},
  \end{equation}
  %------------------------------------
  %------------------------------------
  \begin{equation}
 {\langle}D(p_{3}){\vert}c_{i}(0)\bar{q}_{j}(z){\vert}0{\rangle}\,
 =\, \frac{i\,f_{D}}{4} {\int}dk_{3}\,e^{ik_{3}{\cdot}z}\,
  \Big\{ {\gamma}_{5}\Big[ \!\!\not{p}_{3}\, {\phi}_{D}^{a}(k_{3})
  +m_{3}\,{\phi}_{D}^{p}(k_{3}) \Big] \Big\}_{ji}
  \label{wave-cq-p},
  \end{equation}
  %------------------------------------
  where $f_{\Upsilon}$, $f_{B_{c}^{\ast}}$, $f_{D}$ are
  decay constants;
  wave functions ${\Phi}_{{\Upsilon},B_{c}^{\ast}}^{v,T}$
  and ${\Phi}_{D}^{a}$ are twist-2;
  ${\Phi}_{{\Upsilon},B_{c}^{\ast}}^{t,V}$
  and ${\Phi}_{D}^{p}$ are twist-3.
  The explicit expressions of DAs are \cite{plb752}
  %-----------------------------------------------------
   \begin{equation}
  {\phi}_{\Upsilon}^{v}(x) =
  {\phi}_{\Upsilon}^{T}(x) =
  A_{1}\, x\bar{x}\,
  {\exp}\Big\{ -\frac{m_{b}^{2}}{8\,{\omega}_{1}^{2}\,x\,\bar{x}} \Big\}
   \label{das-upsilon-a},
   \end{equation}
  %-----------------------------------------------------
   \begin{equation}
  {\phi}_{\Upsilon}^{t}(x) =
  A_{2}\, (\bar{x}-x)^{2}\,
  {\exp}\Big\{ -\frac{m_{b}^{2}}{8\,{\omega}_{1}^{2}\,x\,\bar{x}} \Big\}
   \label{das-upsilon-b},
   \end{equation}
  %-----------------------------------------------------
   \begin{equation}
  {\phi}_{\Upsilon}^{V}(x) =
  A_{3}\, \Big\{ 1+(\bar{x}-x)^{2} \Big\}\,
  {\exp}\Big\{ -\frac{m_{b}^{2}}{8\,{\omega}_{1}^{2}\,x\,\bar{x}} \Big\}
   \label{das-upsilon-c},
   \end{equation}
  %-----------------------------------------------------
   \begin{equation}
  {\phi}_{{\Upsilon}(2S)}^{v,t,T,V}(x) = B_{i}\,
  {\phi}_{{\Upsilon}(1S)}^{v,t,T,V}(x)\,
   \Big\{ 1+\frac{m_{b}^{2}}{2\,{\omega}_{1}^{2}\,x\,\bar{x}} \Big\}
   \label{das-upsilon-d},
   \end{equation}
  %-----------------------------------------------------
   \begin{equation}
  {\phi}_{{\Upsilon}(3S)}^{v,t,T,V}(x) = C_{i}\,
  {\phi}_{{\Upsilon}(1S)}^{v,t,T,V}(x)\,
   \Big\{ \Big( 1-\frac{m_{b}^{2}}{2\,{\omega}_{1}^{2}\,x\,\bar{x}} \Big)^{2}
   +6 \Big\}
   \label{das-upsilon-e},
   \end{equation}
  %-----------------------------------------------------
  %-----------------------------------------------------
   \begin{equation}
  {\phi}_{B_{c}^{\ast}}^{v}(x) =
  {\phi}_{B_{c}^{\ast}}^{T}(x) =
  D_{1}\, x\bar{x}\,
  {\exp}\Big\{ -\frac{\bar{x}\,m_{c}^{2}+x\,m_{b}^{2}}
                     {8\,{\omega}_{2}^{2}\,x\,\bar{x}} \Big\}
   \label{das-bc-a},
   \end{equation}
  %-----------------------------------------------------
   \begin{equation}
  {\phi}_{B_{c}^{\ast}}^{t}(x) =
  D_{2}\,  (\bar{x}-x)^{2}\,
  {\exp}\Big\{ -\frac{\bar{x}\,m_{c}^{2}+x\,m_{b}^{2}}
                     {8\,{\omega}_{2}^{2}\,x\,\bar{x}} \Big\}
   \label{das-bc-b},
   \end{equation}
  %-----------------------------------------------------
   \begin{equation}
  {\phi}_{B_{c}^{\ast}}^{V}(x) =
  D_{3}\,  \Big\{ 1+(\bar{x}-x)^{2} \Big\}\,
  {\exp}\Big\{ -\frac{\bar{x}\,m_{c}^{2}+x\,m_{b}^{2}}
                     {8\,{\omega}_{2}^{2}\,x\,\bar{x}} \Big\}
   \label{das-bc-c},
   \end{equation}
  %-----------------------------------------------------
  %-----------------------------------------------------
   \begin{equation}
  {\phi}_{D}^{a}(x) = E_{1}\, x\bar{x}\,
  {\exp}\Big\{ -\frac{\bar{x}\,m_{q}^{2}+x\,m_{c}^{2}}
                     {8\,{\omega}_{3}^{2}\,x\,\bar{x}} \Big\}
   \label{das-cq-a},
   \end{equation}
  %-----------------------------------------------------
   \begin{equation}
  {\phi}_{D}^{p}(x) = E_{2}\,
  {\exp}\Big\{ -\frac{\bar{x}\,m_{q}^{2}+x\,m_{c}^{2}}
                     {8\,{\omega}_{3}^{2}\,x\,\bar{x}} \Big\}
   \label{das-cq-p},
   \end{equation}
  %-----------------------------------------------------
   where $\bar{x}$ $=$ $1$ $-$ $x$; parameter
   ${\omega}_{i}$ $=$ $m_{i}\,{\alpha}_{s}(m_{i})$
   determines the average transverse quark momentum
   according to nonrelativistic quantum chromodynamics
   (NRQCD) power counting rules \cite{prd46};
   parameters $A_{i}$, $B_{i}$, $C_{i}$, $D_{i}$, $E_{i}$ are
   normalization coefficients,
  %-----------------------------------------------------
   \begin{equation}
  {\int}_{0}^{1}dx\,{\phi}_{\Upsilon}^{i}(x) = 1,
   \quad \text{for}\ \ i=v,t,V,T
   \label{normalization-bb},
   \end{equation}
  %-----------------------------------------------------
   \begin{equation}
  {\int}_{0}^{1}dx\,{\phi}_{B_{c}^{\ast}}^{i}(x) = 1,
   \quad \text{for}\ \ i=v,t,V,T
   \label{normalization-bc},
   \end{equation}
  %-----------------------------------------------------
   \begin{equation}
  {\int}_{0}^{1}dx\,{\phi}_{D}^{i}(x) = 1,
   \quad \text{for}\ \ i=a,p
   \label{normalization-cq}.
   \end{equation}
  %-----------------------------------------------------

  %-----------------------------------------------------
  \begin{figure}[h]
  \includegraphics[width=0.98\textwidth,bb=70 530 540 720]{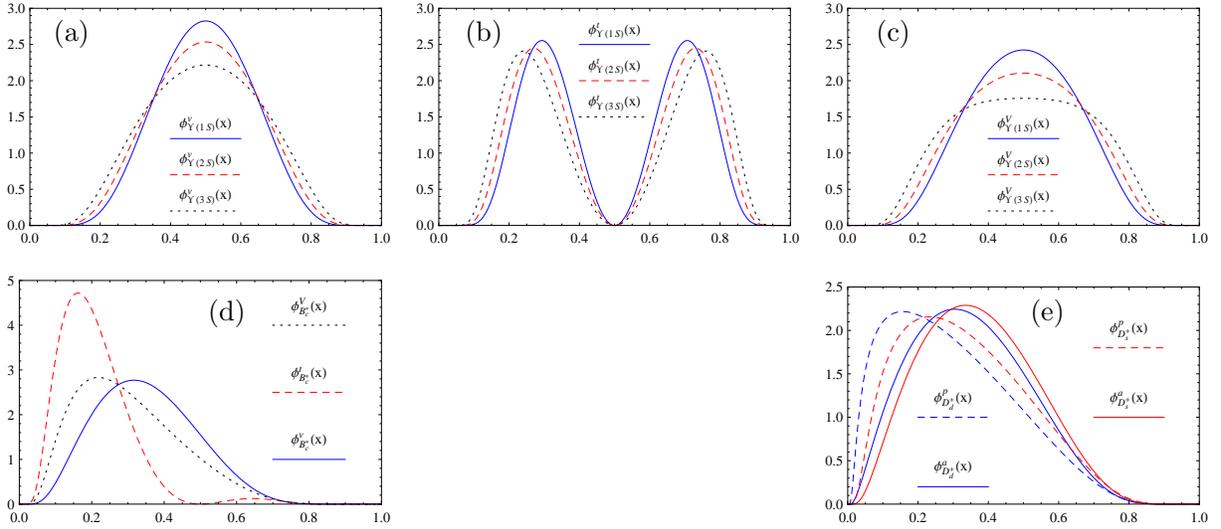}
  \caption{The normalized distribution amplitudes for ${\Upsilon}(nS)$,
  $B_{c}^{\ast}$, $D$ mesons.}
  \label{fig:wave}
  \end{figure}
  %-----------------------------------------------------

  The shape lines of DAs for ${\Upsilon}(nS)$, $B_{c}^{\ast}$, $D$
  mesons  are displayed in Fig.\ref{fig:wave}.
  It is clearly seen that DAs fall quickly down to zero at
  endpoint $x$, $\bar{x}$ ${\to}$ $0$ due to suppression
  from exponential functions, which offer a natural cutoff
  for soft contributions.

  %-----------------------------------------------------
  %-----------------------------------------------------

  %%%%%%%%%%%%%%%%%%%%%%%%%%
  \subsection{Decay amplitudes}
  \label{sec0205}
  The Feynman diagrams for ${\Upsilon}(nS)$ ${\to}$
  $B_{c}^{\ast}D_{s}$ decay are showed in Fig.\ref{fig:fey}.
  There are two types. One is emission topology, and the
  other is annihilation topology. Each type is further
  subdivided into factorizable and nonfactorizable
  diagrams.

  After a detail calculation, amplitude for ${\Upsilon}(nS)$
  ${\to}$ $B_{c}^{\ast}D$ decay is written as
  %-----------------------------------------------------
   \begin{equation}
  {\cal A}({\Upsilon}{\to}B_{c}^{\ast}D)\ =\
  {\cal A}_{L}({\epsilon}_{1}^{{\parallel}},{\epsilon}_{2}^{{\parallel}})
 +{\cal A}_{N}({\epsilon}_{1}^{{\perp}}{\cdot}{\epsilon}_{2}^{{\perp}})
 +i\,{\cal A}_{T}\,{\varepsilon}_{{\mu}{\nu}{\alpha}{\beta}}\,
  {\epsilon}_{1}^{{\mu}}\,{\epsilon}_{2}^{{\nu}}\,
   p_{1}^{\alpha}\,p_{2}^{\beta}
   \label{eq:amp01},
   \end{equation}
  %-----------------------------------------------------
  which is also written as the helicity amplitudes,
  %-----------------------------------------------------
   \begin{equation}
  {\cal M}_{0}\ =\ -{\cal C}\,
  {\cal A}_{L}({\epsilon}_{1}^{{\parallel}},{\epsilon}_{2}^{{\parallel}})
   \label{eq:amp02},
   \end{equation}
  %-----------------------------------------------------
   \begin{equation}
  {\cal M}_{\parallel}\ =\ \sqrt{2}\,{\cal C}\, {\cal A}_{N}
   \label{eq:amp03},
   \end{equation}
  %-----------------------------------------------------
   \begin{equation}
  {\cal M}_{\perp}\ =\ \sqrt{2}\,{\cal C}\,m_{1}\,p\, {\cal A}_{T}
   \label{eq:amp04},
   \end{equation}
  %-----------------------------------------------------
   \begin{equation}
  {\cal C}\ =\ i\frac{G_{F}}{\sqrt{2}}\,\frac{C_{F}}{N_{c}}\,
  {\pi}\, f_{{\Upsilon}}\,f_{B_{c}^{\ast}}\, f_{D}
   \label{eq:amp05},
   \end{equation}
  %-----------------------------------------------------
  where $C_{F}$ $=$ $4/3$ and the color number $N_{c}$ $=$ $3$.

  The expression of polarization amplitude ${\cal A}_{j}$ is
  %-----------------------------------------------------
   \begin{eqnarray}
  {\cal A}_{j} &=&
   V_{cb} V_{cq}^{\ast}\,\Big\{
   \Big( {\cal A}_{a,j}^{LL}+{\cal A}_{b,j}^{LL} \Big) a_{1}
  +\Big( {\cal A}_{c,j}^{LL}+{\cal A}_{d,j}^{LL} \Big) C_{2}
   \Big\}
   \nonumber \\ &-&
   V_{tb} V_{tq}^{\ast}\, \Big\{
   \Big( {\cal A}_{a,j}^{LL}+{\cal A}_{b,j}^{LL} \Big)\,(a_{4}+a_{10})
  +\Big( {\cal A}_{c,j}^{LL}+{\cal A}_{d,j}^{LL} \Big)\, (C_{3}+C_{9})
   \nonumber \\ & & \quad
  +\Big( {\cal A}_{e,j}^{LL}+{\cal A}_{f,j}^{LL} \Big)\,
  (C_{3}+C_{4}-\frac{1}{2}C_{9}-\frac{1}{2}C_{10})
   \nonumber \\ & & \quad
  +\Big( {\cal A}_{g,j}^{LL}+{\cal A}_{h,j}^{LL} \Big)\,
  (a_{3}+a_{4}-\frac{1}{2}a_{9}-\frac{1}{2}a_{10})
   \nonumber \\ & & \quad
  +\Big( {\cal A}_{e,j}^{LR}+{\cal A}_{f,j}^{LR} \Big)\,
   (C_{6}-\frac{1}{2}C_{8})
  +\Big( {\cal A}_{g,j}^{LR}+{\cal A}_{h,j}^{LR} \Big)\,
   (a_{5}-\frac{1}{2}a_{7})
   \nonumber \\ & & \quad
  +\Big( {\cal A}_{c,j}^{SP}+{\cal A}_{d,j}^{SP} \Big)\,
   (C_{5}+C_{7})
  +\Big( {\cal A}_{e,j}^{SP}+{\cal A}_{f,j}^{SP} \Big)\,
   (C_{5}-\frac{1}{2}C_{7}) \Big\}
   \label{amp-all},
   \end{eqnarray}
  %-----------------------------------------------------
  where the subscript $j$ $=$ $L$, $N$, $T$ denotes to three
  different helicity amplitudes; the expressions
  of building blocks ${\cal A}_{i,j}^{k}$ are collected in
  Appendix; $C_{i}$ is Wilson coefficient,
  parameter $a_{i}$ is defined as
  %-----------------------------------------------------
  \begin{equation}
  a_{i} = \left\{ \begin{array}{l}
  C_{i}+C_{i+1}/N_{c}, \quad \text{for odd $i$}; \\
  C_{i}+C_{i-1}/N_{c}, \quad \text{for even $i$}.
  \end{array} \right.
  \label{eq:ai}
  \end{equation}
  %-----------------------------------------------------

  %%%%%%%%%%%%%%%%%%%%%%%%%%%%%%
  \section{Numerical results and discussion}
  \label{sec03}
  In the ${\Upsilon}(nS)$ rest frame, branching ratio for
  ${\Upsilon}(nS)$ ${\to}$ $B_{c}^{\ast}D$ decay is defined as
 %-----------------------------------------------------
   \begin{equation}
  {\cal B}r\ =\ \frac{1}{12{\pi}}\,
   \frac{p}{m_{{\Upsilon}}^{2}\,{\Gamma}_{{\Upsilon}}}\, \Big\{
  {\vert}{\cal M}_{0}{\vert}^{2}+{\vert}{\cal M}_{\parallel}{\vert}^{2}
 +{\vert}{\cal M}_{\perp}{\vert}^{2} \Big\}
   \label{br},
   \end{equation}
 %-----------------------------------------------------
  where $p$ is the center-of-mass momentum of final states;
  ${\Gamma}_{\Upsilon}$ is a total decay width.

  The input parameters are listed in Table \ref{tab:input}.
  If it is not stated explicitly, their central values will
  be used as the default inputs.
  Our numerical results are collected in Table. \ref{tab:br},
  where theoretical uncertainties come from scale $(1{\pm}0.1)t$,
  mass $m_{b}$ and $m_{c}$, and CKM parameters, respectively.
  The following is some comments.

  %---------------------------------------------------------
   \begin{table}[h]
   \caption{The numerical values of input parameters.}
   \label{tab:input}
   \begin{ruledtabular}
   \begin{tabular}{lll}
   \multicolumn{3}{c}{Wolfenstein parameters\footnotemark[1] \cite{pdg}} \\ \hline
   \multicolumn{3}{c}{$A$ $=$ $0.814^{+0.023}_{-0.024}$, \quad
   ${\lambda}$ $=$ $0.22537{\pm}0.00061$, \quad
   $\bar{\rho}$ $=$ $0.117{\pm}0.021$, \quad
   $\bar{\eta}$ $=$ $0.353{\pm}0.013$ } \\ \hline
    \multicolumn{3}{c}{mass, width and decay constant} \\ \hline
    $m_{{\Upsilon}(1S)}$ $=$ $9460.30{\pm}0.26$ MeV \cite{pdg},
  & ${\Gamma}_{{\Upsilon}(1S)}$ $=$ $54.02{\pm}1.25$ keV \cite{pdg},
  & $f_{{\Upsilon}(1S)}$ $=$ $676.4{\pm}10.7$ MeV \cite{plb751}, \\
    $m_{{\Upsilon}(2S)}$ $=$ $10023.26{\pm}0.31$ MeV \cite{pdg},
  & ${\Gamma}_{{\Upsilon}(2S)}$ $=$ $31.98{\pm}2.63$ keV \cite{pdg},
  & $f_{{\Upsilon}(2S)}$ $=$ $473.0{\pm}23.7$ MeV \cite{plb751}, \\
    $m_{{\Upsilon}(3S)}$ $=$ $10355.2{\pm}0.5$ MeV \cite{pdg},
  & ${\Gamma}_{{\Upsilon}(3S)}$ $=$ $20.32{\pm}1.85$ keV \cite{pdg},
  & $f_{{\Upsilon}(3S)}$ $=$ $409.5{\pm}29.4$ MeV \cite{plb751}, \\
    $m_{B_{c}^{\ast}}$ $=$ $6332{\pm}9$ MeV \cite{prd86},
  & $f_{B_{c}^{\ast}}$ $=$ $422{\pm}13$ MeV \cite{prd91},
  & $m_{b}$ $=$ $4.78{\pm}0.06$ GeV \cite{pdg}, \\
    $m_{D_{s}}$ $=$ $1968.30{\pm}0.11$ MeV \cite{pdg},
  & $f_{D_{s}}$ $=$ $257.5{\pm}4.6$ MeV \cite{pdg},
  & $m_{c}$ $=$ $1.67{\pm}0.07$ GeV \cite{pdg}, \\
    $m_{D_{d}}$ $=$ $1869.61{\pm}0.10$ MeV \cite{pdg},
  & $f_{D_{d}}$ $=$ $204.6{\pm}5.0$ MeV \cite{pdg},
  & $m_{s}$ ${\simeq}$ $0.51$ GeV \cite{book}, \\
    ${\Lambda}_{\rm QCD}^{(5)}$ $=$ $214{\pm}7$ MeV \cite{pdg},
  & ${\Lambda}_{\rm QCD}^{(4)}$ $=$ $297{\pm}8$ MeV \cite{pdg},
  & $m_{d}$ ${\simeq}$ $0.31$ GeV \cite{book}.
   \end{tabular}
   \end{ruledtabular}
   \footnotetext[1]{The relation between parameters (${\rho}$, ${\eta}$)
   and ($\bar{\rho}$, $\bar{\eta}$) is \cite{pdg}: $({\rho}+i{\eta})$ $=$
    $\displaystyle \frac{ \sqrt{1-A^{2}{\lambda}^{4}}(\bar{\rho}+i\bar{\eta}) }
    { \sqrt{1-{\lambda}^{2}}[1-A^{2}{\lambda}^{4}(\bar{\rho}+i\bar{\eta})] }$.}
   \end{table}
  %---------------------------------------------------------

  %---------------------------------------------------------
   \begin{table}[h]
   \caption{Branching ratio for ${\Upsilon}(nS)$ ${\to}$ $B_{c}^{\ast}D$ decay.}
   \label{tab:br}
   \begin{ruledtabular}
   \begin{tabular}{cccc}
    & ${\Upsilon}(1S)$ & ${\Upsilon}(2S)$ & ${\Upsilon}(3S)$ \\ \hline
      $10^{10}{\times}{\cal B}r({\Upsilon}{\to}B_{c}^{\ast}D_{s})$
    & $13.21^{+1.58+0.52+0.94}_{-0.75-0.62-0.90}$
    & $ 7.82^{+0.77+0.26+0.56}_{-0.36-0.82-0.53}$
    & $ 7.23^{+0.71+0.47+0.51}_{-0.36-2.31-0.49}$ \\
      $10^{11}{\times}{\cal B}r({\Upsilon}{\to}B_{c}^{\ast}D_{d})$
    & $ 4.54^{+0.50+0.16+0.37}_{-0.24-0.20-0.35}$
    & $ 2.62^{+0.23+0.11+0.21}_{-0.11-0.32-0.20}$
    & $ 2.50^{+0.24+0.19+0.20}_{-0.11-0.85-0.19}$
   \end{tabular}
   \end{ruledtabular}
   \end{table}
  %---------------------------------------------------------

  (1)
  By and large, due to the hierarchical structure of CKM factors
  ${\vert}V_{cb}V_{cs}^{\ast}{\vert}$ $>$ ${\vert}V_{cb}V_{cd}^{\ast}{\vert}$,
  there is a general hierarchical relationship among branching ratios
  ${\cal B}r({\Upsilon}(nS){\to}B_{c}^{\ast}D_{s})$ $>$
  ${\cal B}r({\Upsilon}(nS){\to}B_{c}^{\ast}D_{d})$.

  (2)
  In principle, it is expected to have relations
  ${\cal B}r({\Upsilon}(3S){\to}B_{c}^{\ast}D)$ $>$
  ${\cal B}r({\Upsilon}(2S){\to}B_{c}^{\ast}D)$ $>$
  ${\cal B}r({\Upsilon}(1S){\to}B_{c}^{\ast}D)$ for the
  same $D$ meson, due to the
  fact ${\Gamma}_{{\Upsilon}(3S)}$
  $<$ ${\Gamma}_{{\Upsilon}(2S)}$
  $<$ ${\Gamma}_{{\Upsilon}(1S)}$.
  However, the numbers in Table \ref{tab:br} are beyond our
  expectation.
  Why is it that? Besides convolution integral of DAs resulting
  in different ${\Upsilon}(nS)$ ${\to}$ $B_{c}^{\ast}$ transition
  form factors, one of the possible essential causation is that
  branching ratio is proportional to factor
  $f^{2}_{\Upsilon}/m^{2}_{\Upsilon}\,{\Gamma}_{\Upsilon}$
  and
  %-----------------------------------------------------
   \begin{equation}
   \frac{f^{2}_{{\Upsilon}(1S)}}{m^{2}_{{\Upsilon}(1S)}\,{\Gamma}_{{\Upsilon}(1S)}} :
   \frac{f^{2}_{{\Upsilon}(2S)}}{m^{2}_{{\Upsilon}(2S)}\,{\Gamma}_{{\Upsilon}(2S)}} :
   \frac{f^{2}_{{\Upsilon}(3S)}}{m^{2}_{{\Upsilon}(3S)}\,{\Gamma}_{{\Upsilon}(3S)}}
   \ {\simeq}\ 1.2 : 0.9: 1.0
   \label{eq:bf-phase}.
   \end{equation}
  %-----------------------------------------------------

  (3)
  Branching ratios for ${\Upsilon}(nS)$ ${\to}$ $B_{c}^{\ast}D_{s}$
  decays can reach up to ${\cal O}(10^{-10})$.
  In the center-of-mass frame of ${\Upsilon}(nS)$, the
  final states are back-to-back, and have opposite electric
  charges. In addition to abundant ${\Upsilon}(nS)$ data
  samples in the future experiment, the ``charge tag''
  and ``flavor tag'' technique can be used to effectively
  reconstruct events and reduce background. So
  ${\Upsilon}(nS)$ ${\to}$ $B_{c}^{\ast}D_{s}$ decay might be
  measurable at the running LHC and forthcoming SuperKEKB.
  For example, the ${\Upsilon}(nS)$ production cross section in
  p-Pb collision is about a few ${\mu}b$ at LHCb \cite{jhep1407}
  and ALICE \cite{plb740}. More than $10^{11}$ ${\Upsilon}(nS)$
  data samples per $ab^{-1}$ data collected at LHCb and ALICE
  are in principle available, corresponding to dozens of
  ${\Upsilon}(nS)$ ${\to}$ $B_{c}^{\ast}D_{s}$ events.

  %-----------------------------------------------------
  \begin{figure}[h]
  \includegraphics[width=0.5\textwidth,bb=150 115 460 720]{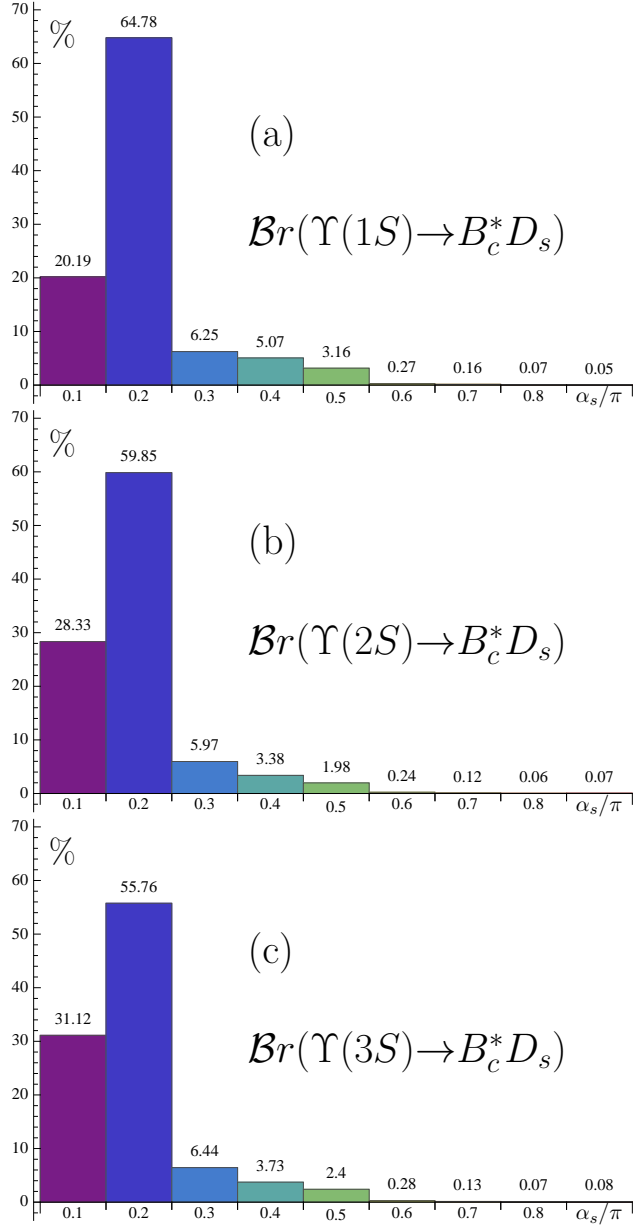}
  \caption{Contributions to branching ratio from different
  regions of ${\alpha}_{s}/{\pi}$ (horizontal axis), where the
  numbers over histogram denote the percentage of the corresponding
  contributions.}
  \label{fig:br-as}
  \end{figure}
  %-----------------------------------------------------

  (4)
  The momentum transition in ${\Upsilon}(nS)$ ${\to}$ $B_{c}^{\ast}D$
  decay may be not large enough, because of $m_{B_{c}^{\ast}}$ $+$
  $m_{D}$ $>$ 8 GeV.
  It is natural to question the validity of perturbative calculation
  with pQCD approach. Therefore, it is very necessary to check what
  percentage of contributions comes from the perturbative region.
  In Fig.\ref{fig:br-as}, contributions to branching ratio from
  different regions of ${\alpha}_{s}/{\pi}$ are plotted.
  It is clearly seen that more than 80\% (90\%) contributions
  come from ${\alpha}_{s}/{\pi}$ ${\le}$ $0.2$ ($0.3$) regions,
  implying that pQCD approach is applicable to the concerned processes,
  and many combined factors (such as the choice of scale $t$, Sudakov
  factor, wave function models, and so on) ensure a reliable
  perturbative calculation.
  Compared with the second bin contribution where
  ${\alpha}_{s}/{\pi}$ $=$ $0.2$, the first bin contribution where
  ${\alpha}_{s}/{\pi}$ $=$ $0.1$ is relatively small.
  One of crucial reasons might be that the absolute values of parameter
  $a_{1}$ and coupling ${\alpha}_{s}$ decrease along with the
  increase of renormalization scale.

  (5)
  Besides the uncertainties listed in Table \ref{tab:br},
  decay constants $f_{\Upsilon}$ and $f_{B_{c}^{\ast}}$
  (decay width ${\Gamma}_{\Upsilon}$) can bring about 7\% (2\%),
  12\% (8\%), 16\% (9\%) uncertainties for ${\Upsilon}(1S)$,
  ${\Upsilon}(2S)$, ${\Upsilon}(3S)$ decays, respectively,
  mainly from $f_{{\Upsilon}(2S,3S)}$ and ${\Gamma}_{{\Upsilon}(2S,3S)}$.
  These are at least two ways to reduce theoretical uncertainty.
  One way is to construct some relative ratios,
  for example, ${\cal B}r({\Upsilon}(nS){\to}B_{c}^{\ast}D_{d})/{\cal B}r({\Upsilon}(nS){\to}B_{c}^{\ast}D_{s})$
  and ${\cal B}r({\Upsilon}(mS){\to}B_{c}^{\ast}D)/{\cal B}r({\Upsilon}(nS){\to}B_{c}^{\ast}D)$.
  The other is to consider higher order corrections to HME,
  more realistic DAs models, and so on.
  Our results are just an order of magnitude
  estimation on branching ratio.

  %%%%%%%%%%%%%%%%%%%%%%%%%%%%%%
  \section{Summary}
  \label{sec04}
  With anticipation of the potential prospects of ${\Upsilon}(nS)$
  physics at high-luminosity heavy-flavor factories,
  search for ${\Upsilon}(nS)$ weak decay seems to be experimentally
  feasible. A theoretical study of ${\Upsilon}(nS)$ weak decay is
  seasonable and necessary.
  In this paper, we investigated the bottom- and charm-changing
  ${\Upsilon}(nS)$ ${\to}$ $B_{c}^{\ast}D_{s,d}$ decays with
  phenomenological pQCD approach.
  It is expected that branching ratio for ${\Upsilon}(nS)$ ${\to}$
  $B_{c}^{\ast}D_{s}$ decay could be up to ${\cal O}(10^{-10})$,
  which might be measurable at the future LHC and SuperKEKB
  experiments.

  %%%%%%%%%%%%%%%%%%%%%%%%%%%%%%
  \section*{Acknowledgments}
  The work is supported by the National Natural Science Foundation
  of China (Grant Nos. 11547014, 11475055, U1332103 and 11275057).

  \begin{appendix}
  %%%%%%%%%%%%%%%%%%%%%%%%%%%%%%
  \section{Building blocks of decay amplitudes}
  \label{blocks}
  For the sake of simplicity,
  we decompose the amplitude Eq.(\ref{amp-all})
  into some building blocks ${\cal A}_{i,j}^{k}$, where
  the subscript $i$ corresponds to the indices of Fig.\ref{fig:fey};
  the subscript $j$ $=$ $L$, $N$, $T$ relates with different
  helicity amplitudes;
  the superscript $k$ refers to one of the three possible Dirac
  structures ${\Gamma}_{1}{\otimes}{\Gamma}_{2}$ of the
  four-quark operator
  $(\bar{q}_{1}{\Gamma}_{1}q_{2})(\bar{q}_{1}{\Gamma}_{2}q_{2})$,
  namely
  $k$ $=$ $LL$ for $(V-A){\otimes}(V-A)$,
  $k$ $=$ $LR$ for $(V-A){\otimes}(V+A)$, and
  $k$ $=$ $SP$ for $-2(S-P){\otimes}(S+P)$.
  The explicit expressions of ${\cal A}_{i,j}^{k}$
  are written as follows.
  %-----------------------------------------------------
  %------------------------------------
   \begin{equation}
  {\cal A}_{a,L}^{LL}\ =\
  {\cal I}_{a}\,
  {\phi}_{\Upsilon}^{v}(x_{1})\,
   \Big\{ {\phi}_{B_{c}^{\ast}}^{v}(x_{2})\, \Big[
   m_{1}^{2}\,s - (4\,m_{1}^{2}\,p^{2}+m_{2}^{2}\,u)\,\bar{x}_{2}
   \Big]+ {\phi}_{B_{c}^{\ast}}^{t}(x_{2})\,m_{2}\,m_{b}\,u \Big\}
   \label{figaL-LL},
   \end{equation}
  %------------------------------------
   \begin{equation}
  {\cal A}_{a,N}^{LL}\ =\
  m_{1}\, {\cal I}_{a}\,
  {\phi}_{\Upsilon}^{V}(x_{1})\,
   \Big\{ {\phi}_{B_{c}^{\ast}}^{V}(x_{2})\,
  m_{2}\, (u-s\,\bar{x}_{2})
 +{\phi}_{B_{c}^{\ast}}^{T}(x_{2})\,m_{b}\,s \Big\}
   \label{figaN-LL},
   \end{equation}
  %------------------------------------
  \begin{equation}
  {\cal A}_{a,T}^{LL}\ =\
  -2\,m_{1}\,
  {\cal I}_{a}\,
  {\phi}_{\Upsilon}^{V}(x_{1})\,
   \Big\{ {\phi}_{B_{c}^{\ast}}^{V}(x_{2})\,m_{2}\,x_{2}
 +{\phi}_{B_{c}^{\ast}}^{T}(x_{2})\,m_{b} \Big\}
   \label{figaT-LL},
   \end{equation}
  %------------------------------------
  %------------------------------------
   \begin{equation}
  {\cal A}_{a,L}^{SP}\ =\
  2\,m_{3}\,
  {\cal I}_{a}\,
  {\phi}_{\Upsilon}^{v}(x_{1})\, \Big\{
  {\phi}_{B_{c}^{\ast}}^{v}(x_{2})\,m_{b}\,t
 +{\phi}_{B_{c}^{\ast}}^{t}(x_{2})\,m_{2}\,
  (2\,m_{1}^{2}-t\,\bar{x}_{2}) \Big\}
   \label{figaL-SP},
   \end{equation}
  %------------------------------------
   \begin{equation}
  {\cal A}_{a,N}^{SP}\ =\
  2\, m_{1}\,m_{3}\,
  {\cal I}_{a}\,
  {\phi}_{\Upsilon}^{V}(x_{1})\, \Big\{
  {\phi}_{B_{c}^{\ast}}^{V}(x_{2})\,2\,m_{2}\,m_{b}
 +{\phi}_{B_{c}^{\ast}}^{T}(x_{2})\,
  (t-2\,m_{2}^{2}\,\bar{x}_{2}) \Big\}
   \label{figaN-SP},
   \end{equation}
  %------------------------------------
  \begin{equation}
  {\cal A}_{a,T}^{SP}\ =\
  4\,m_{1}\,m_{3}\,
  {\cal I}_{a}\,
  {\phi}_{\Upsilon}^{V}(x_{1})\,
  {\phi}_{B_{c}^{\ast}}^{T}(x_{2})
   \label{figaT-SP},
   \end{equation}
  %------------------------------------
  %-----------------------------------------------------
  %------------------------------------
   \begin{equation}
  {\cal A}_{b,L}^{LL}\ =\
  {\cal I}_{b}\,
  {\phi}_{B_{c}^{\ast}}^{v}(x_{2})\,
   \Big\{ {\phi}_{\Upsilon}^{v}(x_{1})\, \Big[
   m_{2}^{2}\,u-m_{1}^{2}\,(s-4\,p^{2})\,\bar{x}_{1} \Big]
 +{\phi}_{\Upsilon}^{t}(x_{1})\, m_{1}\,m_{c}\,s \Big\}
   \label{figbL-LL},
   \end{equation}
  %------------------------------------
   \begin{equation}
  {\cal A}_{b,N}^{LL}\ =\
  m_{2}\, {\cal I}_{b}\,
  {\phi}_{B_{c}^{\ast}}^{V}(x_{2})\,
   \Big\{  {\phi}_{\Upsilon}^{V}(x_{1})\,m_{1}\, (s-u\,\bar{x}_{1})
 +{\phi}_{\Upsilon}^{T}(x_{1})\,m_{c}\,u \Big\}
   \label{figbN-LL},
   \end{equation}
  %------------------------------------
   \begin{equation}
  {\cal A}_{b,T}^{LL}\ =\
  - 2\,m_{2}\, {\cal I}_{b}\,
  {\phi}_{B_{c}^{\ast}}^{V}(x_{2})\, \Big\{
  {\phi}_{\Upsilon}^{V}(x_{1})\,m_{1}\,x_{1}
 +{\phi}_{\Upsilon}^{T}(x_{1})\,m_{c} \Big\}
   \label{figbT-LL},
   \end{equation}
  %------------------------------------
  %------------------------------------
   \begin{equation}
  {\cal A}_{b,L}^{SP}\ =\
  2\,m_{3}\, {\cal I}_{b}\,
  {\phi}_{B_{c}^{\ast}}^{v}(x_{2})\,
   \Big\{ {\phi}_{\Upsilon}^{v}(x_{1})\, m_{c}\,t
 +{\phi}_{\Upsilon}^{t}(x_{1})\, m_{1}\,
  (2\,m_{2}^{2}-t\,\bar{x}_{1}) \Big\}
   \label{figbL-SP},
   \end{equation}
  %------------------------------------
   \begin{equation}
  {\cal A}_{b,N}^{SP}\ =\
  2\,m_{2}\, m_{3}\, {\cal I}_{b}\,
  {\phi}_{B_{c}^{\ast}}^{V}(x_{2})\,\Big\{
  {\phi}_{\Upsilon}^{V}(x_{1})\,2\,m_{1}\,m_{c}
 +{\phi}_{\Upsilon}^{T}(x_{1})\,
  (t-2\,m_{1}^{2}\,\bar{x}_{1}) \Big\}
   \label{figbN-SP},
   \end{equation}
  %------------------------------------
   \begin{equation}
  {\cal A}_{b,T}^{SP}\ =\
  4\,m_{2}\,m_{3}\,{\cal I}_{b}\,
  {\phi}_{\Upsilon}^{T}(x_{1})\,
  {\phi}_{B_{c}^{\ast}}^{V}(x_{2})
   \label{figbT-SP},
   \end{equation}
  %------------------------------------
  %-----------------------------------------------------
  %------------------------------------
   \begin{eqnarray}
  {\cal A}_{c,L}^{LL} &=&
  {\cal I}_{c}\,
  {\phi}_{D}^{a}(x_{3})\, \Big\{
  {\phi}_{\Upsilon}^{v}(x_{1})\,
  {\phi}_{B_{c}^{\ast}}^{v}(x_{2})\,
  4\,m_{1}^{2}\,p^{2}\,(x_{1}-\bar{x}_{3})
   \nonumber \\ & & +
  {\phi}_{\Upsilon}^{t}(x_{1})\,
  {\phi}_{B_{c}^{\ast}}^{t}(x_{2})\,
  m_{1}\,m_{2}\,
  (u\,x_{1}-s\,x_{2}-2\,m_{3}^{2}\,\bar{x}_{3}) \Big\}
   \label{figcL-LL},
   \end{eqnarray}
  %------------------------------------
   \begin{equation}
  {\cal A}_{c,N}^{LL}\ =\
  {\cal I}_{c}\,
  {\phi}_{\Upsilon}^{T}(x_{1})\,
  {\phi}_{B_{c}^{\ast}}^{T}(x_{2})\,
  {\phi}_{D}^{a}(x_{3})\, \Big\{
  m_{1}^{2}\,s\,(x_{1}-\bar{x}_{3})
   +m_{2}^{2}\,u\,(\bar{x}_{3}-x_{2}) \Big\}
   \label{figcN-LL},
   \end{equation}
  %------------------------------------
   \begin{equation}
  {\cal A}_{c,T}^{LL} \ =\
  2\,{\cal I}_{c}\,
  {\phi}_{\Upsilon}^{T}(x_{1})\,
  {\phi}_{B_{c}^{\ast}}^{T}(x_{2})\,
  {\phi}_{D}^{a}(x_{3})\, \Big\{
   m_{1}^{2}\,(\bar{x}_{3}-x_{1})
  +m_{2}^{2}\,(x_{2}-\bar{x}_{3}) \Big\}
   \label{figcT-LL},
   \end{equation}
  %------------------------------------
  %------------------------------------
   \begin{eqnarray}
  {\cal A}_{c,L}^{SP} &=&
  {\cal I}_{c}\,
  {\phi}_{D}^{p}(x_{3})\,
  \Big\{ {\phi}_{\Upsilon}^{v}(x_{1})\,
  {\phi}_{B_{c}^{\ast}}^{t}(x_{2})\, m_{2}\,m_{3}\,
  (2\,m_{1}^{2}\,x_{1}-t\,x_{2}-u\,\bar{x}_{3})
   \nonumber \\ & &
 +{\phi}_{\Upsilon}^{t}(x_{1})\,
  {\phi}_{B_{c}^{\ast}}^{v}(x_{2})\, m_{1}\,m_{3}\,
  (t\,x_{1}-2\,m_{2}^{2}\,x_{2}-s\,\bar{x}_{3}) \Big\}
   \label{figcL-SP-d},
   \end{eqnarray}
  %------------------------------------
   \begin{eqnarray}
  {\cal A}_{c,N}^{SP} &=&
  {\cal I}_{c}\,
  {\phi}_{D}^{p}(x_{3})\,
   \Big\{ {\phi}_{\Upsilon}^{V}(x_{1})\,
  {\phi}_{B_{c}^{\ast}}^{T}(x_{2})\, m_{1}\,m_{3}\,
  (t\,x_{1}-2\,m_{2}^{2}\,x_{2}-s\,\bar{x}_{3})
   \nonumber \\ & &
 +{\phi}_{\Upsilon}^{T}(x_{1})\,
  {\phi}_{B_{c}^{\ast}}^{V}(x_{2})\, m_{2}\,m_{3}\,
  (2\,m_{1}^{2}\,x_{1}-t\,x_{2}-u\,\bar{x}_{3}) \Big\}
   \label{figcN-SP-d},
   \end{eqnarray}
  %------------------------------------
   \begin{equation}
  {\cal A}_{c,T}^{SP}\ =\
  2\,m_{3}\,{\cal I}_{c}\,
  {\phi}_{D}^{p}(x_{3})\,
  \Big\{ {\phi}_{\Upsilon}^{V}(x_{1})\,
  {\phi}_{B_{c}^{\ast}}^{T}(x_{2})\, m_{1}\, (x_{1}-\bar{x}_{3})
 +{\phi}_{\Upsilon}^{T}(x_{1})\,
  {\phi}_{B_{c}^{\ast}}^{V}(x_{2})\, m_{2}\, (\bar{x}_{3}-x_{2}) \Big\}
   \label{figcT-SP-d},
   \end{equation}
  %------------------------------------
  %-----------------------------------------------------
  %------------------------------------
   \begin{eqnarray}
  {\cal A}_{d,L}^{LL} &=&
  {\cal I}_{d}\,
   \Big\{ {\phi}_{\Upsilon}^{t}(x_{1})\,
  {\phi}_{B_{c}^{\ast}}^{t}(x_{2})\,
  {\phi}_{D}^{a}(x_{3})\,m_{1}\, m_{2}\,
  (s\,x_{2} +2\,m_{3}^{2}\,x_{3}-u\,x_{1})
   \nonumber \\ &+&
  {\phi}_{\Upsilon}^{v}(x_{1})
  {\phi}_{B_{c}^{\ast}}^{v}(x_{2}) \Big[
  {\phi}_{D}^{a}(x_{3})\, 4\,m_{1}^{2}\,p^{2}\, (x_{3}-x_{2})
 -{\phi}_{D}^{p}(x_{3})\, m_{3}\, m_{c}\,t \Big] \Big\}
   \label{figdL-LL-d},
   \end{eqnarray}
  %------------------------------------
   \begin{eqnarray}
  {\cal A}_{d,N}^{LL} &=&
  {\cal I}_{d}\, \Big\{
  {\phi}_{\Upsilon}^{T}(x_{1})\,
  {\phi}_{B_{c}^{\ast}}^{T}(x_{2})\,
  {\phi}_{D}^{a}(x_{3})\, \Big[
   m_{1}^{2}\,s\,(x_{3}-x_{1})
  +m_{2}^{2}\,u\,(x_{2}-x_{3}) \Big]
   \nonumber \\ &-&
  {\phi}_{\Upsilon}^{V}(x_{1})\,
  {\phi}_{B_{c}^{\ast}}^{V}(x_{2})\,
  {\phi}_{D}^{p}(x_{3})\,2\,m_{1}\,m_{2}\,m_{3}\,m_{c} \Big\}
   \label{figdN-LL-d},
   \end{eqnarray}
  %------------------------------------
   \begin{equation}
  {\cal A}_{d,T}^{LL}\ =\
  2\, {\cal I}_{d}\,
  {\phi}_{\Upsilon}^{T}(x_{1})\,
  {\phi}_{B_{c}^{\ast}}^{T}(x_{2})\,
  {\phi}_{D}^{a}(x_{3})\, \Big\{
   m_{1}^{2}\,(x_{1}-x_{3})
  -m_{2}^{2}\,(x_{2}-x_{3}) \Big\}
   \label{figdT-LL-d},
   \end{equation}
  %------------------------------------
   \begin{eqnarray}
  {\cal A}_{d,L}^{SP} &=&
  {\cal I}_{d}\, \Big\{
  {\phi}_{\Upsilon}^{v}(x_{1})\,
  {\phi}_{B_{c}^{\ast}}^{t}(x_{2})\, m_{2}\, \Big[
  {\phi}_{D}^{p}(x_{3})\,m_{3}\,(t\,x_{2}+u\,x_{3} -2\,m_{1}^{2}\,x_{1})
 -{\phi}_{D}^{a}(x_{3})\,m_{c}\,u \Big]
   \nonumber \\ &+&
  {\phi}_{\Upsilon}^{t}(x_{1})\,
  {\phi}_{B_{c}^{\ast}}^{v}(x_{2})\, m_{1}\, \Big[
  {\phi}_{D}^{p}(x_{3})\,m_{3}\,
  (2\,m_{2}^{2}\,x_{2}+s\,x_{3}-t\,x_{1})
 -{\phi}_{D}^{a}(x_{3})\,m_{c}\,s \Big] \Big\}
   \label{figdL-SP-d},
   \end{eqnarray}
  %------------------------------------
   \begin{eqnarray}
  {\cal A}_{d,N}^{SP} &=&
  {\cal I}_{d}\, \Big\{
  {\phi}_{\Upsilon}^{V}(x_{1})\,
  {\phi}_{B_{c}^{\ast}}^{T}(x_{2})\, m_{1}\, \Big[
  {\phi}_{D}^{p}(x_{3})\,m_{3}\,
  (2\,m_{2}^{2}\,x_{2}+s\,x_{3}-t\,x_{1})
 -{\phi}_{D}^{a}(x_{3})\,m_{c}\,s \Big]
   \nonumber \\ &+&
  {\phi}_{\Upsilon}^{T}(x_{1})\,
  {\phi}_{B_{c}^{\ast}}^{V}(x_{2})\, m_{2}\,\Big[
  {\phi}_{D}^{p}(x_{3})\, m_{3}\,
  (t\,x_{2}+u\,x_{3}-2\,m_{1}^{2}\,x_{1})
 -{\phi}_{D}^{a}(x_{3})\,m_{c}\,u \Big] \Big\}
   \label{figdN-SP-d},
   \end{eqnarray}
  %------------------------------------
   \begin{eqnarray}
  {\cal A}_{d,T}^{SP} &=&
  2\, {\cal I}_{d}\, \Big\{
  {\phi}_{\Upsilon}^{V}(x_{1})\,
  {\phi}_{B_{c}^{\ast}}^{T}(x_{2})\, m_{1}\, \Big[
  {\phi}_{D}^{p}(x_{3})\, m_{3}\,(x_{3}-x_{1})
 -{\phi}_{D}^{a}(x_{3})\,m_{c} \Big]
   \nonumber \\ &+&
  {\phi}_{\Upsilon}^{T}(x_{1})\,
  {\phi}_{B_{c}^{\ast}}^{V}(x_{2})\, m_{2}\,\Big[
  {\phi}_{D}^{a}(x_{3})\,m_{c}
 +{\phi}_{D}^{p}(x_{3})\,{m}_{3}\,(x_{2}-x_{3})
    \Big] \Big\}
   \label{figdT-SP-d},
   \end{eqnarray}
  %------------------------------------
  %-----------------------------------------------------
  %------------------------------------
   \begin{eqnarray}
  {\cal A}_{e,L}^{LL} &=&
  {\cal I}_{e}\, \Big\{
  {\phi}_{B_{c}^{\ast}}^{v}(x_{2})\,
  {\phi}_{D}^{a}(x_{3})\, \Big[
  {\phi}_{\Upsilon}^{v}(x_{1})\,4\,m_{1}^{2}\,p^{2}\,(x_{1}-\bar{x}_{3})
 -{\phi}_{\Upsilon}^{t}(x_{1})\,m_{1}\,m_{b}\,s \Big]
   \nonumber \\ &+&
  {\phi}_{\Upsilon}^{v}(x_{1})\,
  {\phi}_{B_{c}^{\ast}}^{t}(x_{2})\,
  {\phi}_{D}^{p}(x_{3})\,m_{2}\,m_{3}\,
  (2\,m_{1}^{2}\,x_{1}-t\,x_{2}-u\,\bar{x}_{3} ) \Big\}
   \label{figeL-LL},
   \end{eqnarray}
  %------------------------------------
   \begin{eqnarray}
  {\cal A}_{e,N}^{LL} &=&
  {\cal I}_{e}\, \Big\{
  {\phi}_{\Upsilon}^{V}(x_{1})
  {\phi}_{B_{c}^{\ast}}^{T}(x_{2})\,
  {\phi}_{D}^{p}(x_{3})\,m_{1}\,m_{3}\,
  (t\,x_{1}-2\,m_{2}^{2}\,x_{2}-s\,\bar{x}_{3})
   \nonumber \\ &-&
  {\phi}_{\Upsilon}^{T}(x_{1})\,
  {\phi}_{B_{c}^{\ast}}^{V}(x_{2})\,
  {\phi}_{D}^{a}(x_{3})\,m_{2}\,m_{b}\,u \Big\}
   \label{figeN-LL},
   \end{eqnarray}
  %------------------------------------
   \begin{equation}
  {\cal A}_{e,T}^{LL}\ =\
  2\, {\cal I}_{e}\, \Big\{
  {\phi}_{\Upsilon}^{V}(x_{1})
  {\phi}_{B_{c}^{\ast}}^{T}(x_{2})\,
  {\phi}_{D}^{p}(x_{3})\,m_{1}\,m_{3}\,
  (x_{1}-\bar{x}_{3})
 -{\phi}_{\Upsilon}^{T}(x_{1})\,
  {\phi}_{B_{c}^{\ast}}^{V}(x_{2})\,
  {\phi}_{D}^{a}(x_{3})\,m_{2}\,m_{b} \Big\}
   \label{figeT-LL},
   \end{equation}
  %------------------------------------
   \begin{eqnarray}
  {\cal A}_{e,L}^{LR} &=&
  {\cal I}_{e}\, \Big\{
  {\phi}_{B_{c}^{\ast}}^{v}(x_{2})\,
  {\phi}_{D}^{a}(x_{3})\, \Big[
  {\phi}_{\Upsilon}^{v}(x_{1})\,t\,
  (s\,x_{2}+2\,m_{3}^{2}\,\bar{x}_{3}-u\,x_{1})
 +{\phi}_{\Upsilon}^{t}(x_{1})\,m_{1}\,m_{b}\,s \Big]
   \nonumber \\ &+&
  {\phi}_{B_{c}^{\ast}}^{t}(x_{2})\,
  {\phi}_{D}^{p}(x_{3})\,m_{2}\,m_{3}\, \Big[
  {\phi}_{\Upsilon}^{v}(x_{1})\,
  (t\,x_{2}+u\,\bar{x}_{3}-2\,m_{1}^{2}\,x_{1})
 +{\phi}_{\Upsilon}^{t}(x_{1})\,4\,m_{1}\,m_{b}
   \Big] \Big\}
   \label{figeL-LR},
   \end{eqnarray}
  %------------------------------------
   \begin{eqnarray}
  {\cal A}_{e,N}^{LR} &=&
  {\cal I}_{e}\, \Big\{
  {\phi}_{B_{c}^{\ast}}^{V}(x_{2})\,
  {\phi}_{D}^{a}(x_{3})\, m_{2}\,\Big[
  {\phi}_{\Upsilon}^{V}(x_{1})\,2\,m_{1}\,
  (s\,x_{2}+2\,m_{3}^{2}\,\bar{x}_{3}-u\,x_{1})
 +{\phi}_{\Upsilon}^{T}(x_{1})\,m_{b}\,u \Big]
   \nonumber \\ &+&
  {\phi}_{B_{c}^{\ast}}^{T}(x_{2})\,
  {\phi}_{D}^{p}(x_{3})\,m_{3}\, \Big[
  {\phi}_{\Upsilon}^{V}(x_{1})\,m_{1}\,
  (2\,m_{2}^{2}\,x_{2}+s\,\bar{x}_{3}-t\,x_{1})
 +{\phi}_{\Upsilon}^{T}(x_{1})\,2\,m_{b}\,t
   \Big] \Big\}
   \label{figeN-LR},
   \end{eqnarray}
  %------------------------------------
   \begin{eqnarray}
  {\cal A}_{e,T}^{LR} &=&
  2\, {\cal I}_{e}\, \Big\{
  {\phi}_{B_{c}^{\ast}}^{T}(x_{2})\,
  {\phi}_{D}^{p}(x_{3})\, m_{3}\,\Big[
  {\phi}_{\Upsilon}^{V}(x_{1})\,m_{1}\,(\bar{x}_{3}-x_{1})
 +{\phi}_{\Upsilon}^{T}(x_{1})\,2\,m_{b} \Big]
   \nonumber \\ &-&
  {\phi}_{\Upsilon}^{T}(x_{1})\,
  {\phi}_{B_{c}^{\ast}}^{V}(x_{2})\,
  {\phi}_{D}^{a}(x_{3})\,m_{2}\,m_{b} \Big\}
   \label{figeT-LR},
   \end{eqnarray}
  %------------------------------------
   \begin{eqnarray}
  {\cal A}_{e,L}^{SP} &=&
  {\cal I}_{e}\, \Big\{
  {\phi}_{B_{c}^{\ast}}^{t}(x_{2})\,
  {\phi}_{D}^{a}(x_{3})\, m_{2}\, \Big[
  {\phi}_{\Upsilon}^{v}(x_{1})\,m_{b}\,u
 +{\phi}_{\Upsilon}^{t}(x_{1})\,m_{1}\,
  (s\,x_{2}+2\,m_{3}^{2}\,\bar{x}_{3}-u\,x_{1}) \Big]
   \nonumber \\ &+&
  {\phi}_{B_{c}^{\ast}}^{v}(x_{2})\,
  {\phi}_{D}^{p}(x_{3})\,m_{3}\, \Big[
  {\phi}_{\Upsilon}^{v}(x_{1})\,m_{b}\,t
 +{\phi}_{\Upsilon}^{t}(x_{1})\,m_{1}\,
  (2\,m_{2}^{2}\,x_{2}+s\,\bar{x}_{3}-t\,x_{1})
   \Big] \Big\}
   \label{figeL-SP},
   \end{eqnarray}
  %------------------------------------
   \begin{eqnarray}
  {\cal A}_{e,N}^{SP} &=&
  {\cal I}_{e}\, \Big\{
  {\phi}_{B_{c}^{\ast}}^{T}(x_{2})\,
  {\phi}_{D}^{a}(x_{3})\, \Big[
  {\phi}_{\Upsilon}^{V}(x_{1})\,m_{1}\,m_{b}\,s
 +{\phi}_{\Upsilon}^{T}(x_{1})\,
  \{ m_{1}^{2}\,s\,(\bar{x}_{3}-x_{1})
   + m_{2}^{2}\,u\,(x_{2}-\bar{x}_{3}) \} \Big]
   \nonumber \\ &+&
  {\phi}_{B_{c}^{\ast}}^{V}(x_{2})\,
  {\phi}_{D}^{p}(x_{3})\,m_{2}\,m_{3}\, \Big[
  {\phi}_{\Upsilon}^{V}(x_{1})\,2\,m_{1}\,m_{b}
 +{\phi}_{\Upsilon}^{T}(x_{1})\,
  (t\,x_{2}+u\,\bar{x}_{3}-2\,m_{1}^{2}\,x_{1})
   \Big] \Big\}
   \label{figeN-SP},
   \end{eqnarray}
  %------------------------------------
   \begin{eqnarray}
  {\cal A}_{e,T}^{SP} &=&
  2\, {\cal I}_{e}\, \Big\{
  {\phi}_{B_{c}^{\ast}}^{T}(x_{2})\,
  {\phi}_{D}^{a}(x_{3})\, \Big[
  {\phi}_{\Upsilon}^{V}(x_{1})\,m_{1}\,m_{b}
 +{\phi}_{\Upsilon}^{T}(x_{1})\,
  \{ m_{1}^{2}\,(\bar{x}_{3}-x_{1})
   + m_{2}^{2}\,(x_{2}-\bar{x}_{3}) \} \Big]
  \nonumber \\ &+&
  {\phi}_{\Upsilon}^{T}(x_{1})\,
  {\phi}_{B_{c}^{\ast}}^{V}(x_{2})\,
  {\phi}_{D}^{p}(x_{3})\,m_{2}\,m_{3}\,(\bar{x}_{3}-x_{2})
   \Big] \Big\}
   \label{figeT-SP},
   \end{eqnarray}
  %------------------------------------
  %-----------------------------------------------------
  %------------------------------------
   \begin{eqnarray}
  {\cal A}_{f,L}^{LL} &=&
  {\cal I}_{f}\, \Big\{
  {\phi}_{B_{c}^{\ast}}^{v}(x_{2})\,
  {\phi}_{D}^{a}(x_{3})\, \Big[
  {\phi}_{\Upsilon}^{v}(x_{1})\,t\,
  (u\,\bar{x}_{1}-s\,x_{2}-2\,m_{3}^{2}\,\bar{x}_{3})
 -{\phi}_{\Upsilon}^{t}(x_{1})\,m_{1}\,m_{b}\,s \Big]
   \nonumber \\ &+&
  {\phi}_{B_{c}^{\ast}}^{t}(x_{2})\,
  {\phi}_{D}^{p}(x_{3})\,m_{2}\,m_{3}\,\Big[
  {\phi}_{\Upsilon}^{v}(x_{1})\,
  (2\,m_{1}^{2}\,\bar{x}_{1}-t\,x_{2}-u\,\bar{x}_{3})
 -{\phi}_{\Upsilon}^{t}(x_{1})\,4\,m_{1}\,m_{b}
   \Big] \Big\}
   \label{figfL-LL},
   \end{eqnarray}
  %------------------------------------
   \begin{eqnarray}
  {\cal A}_{f,N}^{LL} &=&
  {\cal I}_{f}\, \Big\{
  {\phi}_{B_{c}^{\ast}}^{V}(x_{2})\,
  {\phi}_{D}^{a}(x_{3})\, m_{2}\,\Big[
  {\phi}_{\Upsilon}^{V}(x_{1})\,2\,m_{1}\,
  (u\,\bar{x}_{1}-s\,x_{2}-2\,m_{3}^{2}\,\bar{x}_{3})
 -{\phi}_{\Upsilon}^{T}(x_{1})\,m_{b}\,u \Big]
   \nonumber \\ &+&
  {\phi}_{B_{c}^{\ast}}^{T}(x_{2})\,
  {\phi}_{D}^{p}(x_{3})\,m_{3}\,\Big[
  {\phi}_{\Upsilon}^{V}(x_{1})\,m_{1}\,
  (t\,\bar{x}_{1}-2\,m_{2}^{2}\,x_{2}-s\,\bar{x}_{3})
 -{\phi}_{\Upsilon}^{T}(x_{1})\,2\,m_{b}\,t  \Big] \Big\}
   \label{figfN-LL},
   \end{eqnarray}
  %------------------------------------
   \begin{eqnarray}
  {\cal A}_{f,T}^{LL} &=&
  2\, {\cal I}_{f}\, \Big\{
  {\phi}_{B_{c}^{\ast}}^{T}(x_{2})\,
  {\phi}_{D}^{p}(x_{3})\,m_{3}\, \Big[
  {\phi}_{\Upsilon}^{V}(x_{1})\,m_{1}\,(\bar{x}_{1}-\bar{x}_{3})
 -{\phi}_{\Upsilon}^{T}(x_{1})\,2\,m_{b} \Big]
   \nonumber \\ &+&
  {\phi}_{\Upsilon}^{T}(x_{1})\,
  {\phi}_{B_{c}^{\ast}}^{V}(x_{2})\,
  {\phi}_{D}^{a}(x_{3})\,m_{2}\, m_{b}  \Big\}
   \label{figfT-LL},
   \end{eqnarray}
  %------------------------------------
   \begin{eqnarray}
  {\cal A}_{f,L}^{LR} &=&
  {\cal I}_{f}\, \Big\{
  {\phi}_{B_{c}^{\ast}}^{v}(x_{2})\,
  {\phi}_{D}^{a}(x_{3})\, \Big[
  {\phi}_{\Upsilon}^{v}(x_{1})\,4\,m_{1}^{2}\,p^{2}\, (\bar{x}_{3}-\bar{x}_{1})
 +{\phi}_{\Upsilon}^{t}(x_{1})\,m_{1}\,m_{b}\,s \Big]
   \nonumber \\ &+&
  {\phi}_{\Upsilon}^{v}(x_{1})\,
  {\phi}_{B_{c}^{\ast}}^{t}(x_{2})\,
  {\phi}_{D}^{p}(x_{3})\,m_{2}\,m_{3}\,
  (t\,x_{2}+u\,\bar{x}_{3}-2\,m_{1}^{2}\,\bar{x}_{1}) \Big\}
   \label{figfL-LR},
   \end{eqnarray}
  %------------------------------------
   \begin{eqnarray}
  {\cal A}_{f,N}^{LR} &=&
  {\cal I}_{f}\, \Big\{
  {\phi}_{\Upsilon}^{V}(x_{1})\,
  {\phi}_{B_{c}^{\ast}}^{T}(x_{2})\,
  {\phi}_{D}^{p}(x_{3})\,m_{1}\,m_{3}\,
  (2\,m_{2}^{2}\,x_{2}+s\,\bar{x}_{3}-t\,\bar{x}_{1})
   \nonumber \\ &+&
  {\phi}_{\Upsilon}^{T}(x_{1})\,
  {\phi}_{B_{c}^{\ast}}^{V}(x_{2})\,
  {\phi}_{D}^{a}(x_{3})\, m_{2}\,m_{b}\,u \Big\}
   \label{figfN-LR},
   \end{eqnarray}
  %------------------------------------
   \begin{equation}
  {\cal A}_{f,T}^{LR}\ =\
  2\, {\cal I}_{f}\, \Big\{
  {\phi}_{\Upsilon}^{T}(x_{1})\,
  {\phi}_{B_{c}^{\ast}}^{V}(x_{2})\,
  {\phi}_{D}^{a}(x_{3})\, m_{2}\,m_{b}
 +{\phi}_{\Upsilon}^{V}(x_{1})\,
  {\phi}_{B_{c}^{\ast}}^{T}(x_{2})\,
  {\phi}_{D}^{p}(x_{3})\,m_{1}\,m_{3}\,
  (\bar{x}_{3}-\bar{x}_{1}) \Big\}
   \label{figfT-LR},
   \end{equation}
  %------------------------------------
   \begin{eqnarray}
  {\cal A}_{f,L}^{SP} &=&
  {\cal I}_{f}\, \Big\{
  {\phi}_{B_{c}^{\ast}}^{t}(x_{2})\,
  {\phi}_{D}^{a}(x_{3})\,m_{2}\,\Big[
  {\phi}_{\Upsilon}^{v}(x_{1})\,m_{b}\,u
 +{\phi}_{\Upsilon}^{t}(x_{1})\,m_{1}\,
  (s\,x_{2}+2\,m_{3}^{2}\,\bar{x}_{3}-u\,\bar{x}_{1}) \Big]
   \nonumber \\ &+&
  {\phi}_{B_{c}^{\ast}}^{v}(x_{2})\,
  {\phi}_{D}^{p}(x_{3})\,m_{3}\, \Big[
  {\phi}_{\Upsilon}^{v}(x_{1})\,m_{b}\,t
 +{\phi}_{\Upsilon}^{t}(x_{1})\,m_{1}\,
  (2\,m_{2}^{2}\,x_{2}+s\,\bar{x}_{3}-t\,\bar{x}_{1}) \Big] \Big\}
   \label{figfL-SP},
   \end{eqnarray}
  %------------------------------------
   \begin{eqnarray}
  {\cal A}_{f,N}^{SP} &=&
  {\cal I}_{f}\, \Big\{
  {\phi}_{B_{c}^{\ast}}^{T}(x_{2})\,
  {\phi}_{D}^{a}(x_{3})\,\Big[
  {\phi}_{\Upsilon}^{V}(x_{1})\,m_{1}\,m_{b}\,s
 +{\phi}_{\Upsilon}^{T}(x_{1})\,\{
   m_{1}^{2}\,s\,(\bar{x}_{3}-\bar{x}_{1})
  +m_{2}^{2}\,u\,(x_{2}-\bar{x}_{3}) \} \Big]
   \nonumber \\ &+&
  {\phi}_{B_{c}^{\ast}}^{V}(x_{2})\,
  {\phi}_{D}^{p}(x_{3})\,m_{2}\,m_{3}\, \Big[
  {\phi}_{\Upsilon}^{V}(x_{1})\,2\,m_{1}\,m_{b}
 +{\phi}_{\Upsilon}^{T}(x_{1})\,
  (t\,x_{2}+u\,\bar{x}_{3}-2\,m_{1}^{2}\,\bar{x}_{1}) \Big] \Big\}
   \label{figfN-SP},
   \end{eqnarray}
  %------------------------------------
   \begin{eqnarray}
  {\cal A}_{f,T}^{SP} &=&
  2\, {\cal I}_{f}\, \Big\{
  {\phi}_{B_{c}^{\ast}}^{T}(x_{2})\,
  {\phi}_{D}^{a}(x_{3})\,\Big[
  {\phi}_{\Upsilon}^{V}(x_{1})\,m_{1}\,m_{b}
 +{\phi}_{\Upsilon}^{T}(x_{1})\,\{
   m_{1}^{2}\,(\bar{x}_{3}-\bar{x}_{1})
  +m_{2}^{2}\,(x_{2}-\bar{x}_{3}) \} \Big]
   \nonumber \\ &+&
  {\phi}_{\Upsilon}^{T}(x_{1})\,
  {\phi}_{B_{c}^{\ast}}^{V}(x_{2})\,
  {\phi}_{D}^{p}(x_{3})\,m_{2}\,m_{3}\,(\bar{x}_{3}-x_{2})
   \Big\}
   \label{figfT-SP},
   \end{eqnarray}
  %------------------------------------
  %-----------------------------------------------------
  %------------------------------------
   \begin{equation}
  {\cal A}_{g,L}^{LL,LR}=
  {\cal I}_{g}\, \Big\{
  {\phi}_{B_{c}^{\ast}}^{v}(x_{2})\,
  {\phi}_{D}^{a}(x_{3})\,
  ( m_{3}^{2}\,t\,\bar{x}_{2}+ m_{1}^{2}\,s\,x_{2} )
 +{\phi}_{B_{c}^{\ast}}^{t}(x_{2})\,
  {\phi}_{D}^{p}(x_{3})\,2\,m_{2}\,m_{3}\,
  (u+t\,x_{2}) \Big\}
   \label{figgL-LL},
   \end{equation}
  %------------------------------------
   \begin{equation}\!
  {\cal A}_{g,N}^{LL,LR} =
  m_{1}\, {\cal I}_{g}\, \Big\{
  {\phi}_{B_{c}^{\ast}}^{V}(x_{2})\,
  {\phi}_{D}^{a}(x_{3})\,m_{2}\,
  (2\,m_{3}^{2}+s\,x_{2})
 +{\phi}_{B_{c}^{\ast}}^{T}(x_{2})\,
  {\phi}_{D}^{p}(x_{3})\,2\,m_{3}\,
  (s+2\,m_{2}^{2}\,x_{2}) \Big\}
   \label{figgN-LL},
   \end{equation}
  %------------------------------------
   \begin{equation}
  {\cal A}_{g,T}^{LL,LR}=
  2\,m_{1}\, {\cal I}_{g}\,
   \Big\{ {\phi}_{B_{c}^{\ast}}^{T}(x_{2})\,
  {\phi}_{D}^{p}(x_{3})\,2\,m_{3}
 -{\phi}_{B_{c}^{\ast}}^{V}(x_{2})\,
  {\phi}_{D_{q}}^{a}(x_{3})\, m_{2}\,x_{2} \Big\}
   \label{figgT-LL},
   \end{equation}
  %------------------------------------
  %-----------------------------------------------------
  %------------------------------------
   \begin{equation}
  {\cal A}_{h,L}^{LL,LR}\, =\,
  {\cal I}_{h}\,
  {\phi}_{B_{c}^{\ast}}^{v}(x_{2})\, \Big\{
  {\phi}_{D}^{a}(x_{3})\,
  ( m_{2}^{2}\,u\,x_{3}+ m_{1}^{2}\,s\,\bar{x}_{3} )
  -{\phi}_{D}^{p}(x_{3})\,m_{3}\,m_{b}\,t \Big\}
   \label{fighL-LL},
   \end{equation}
  %------------------------------------
   \begin{equation}
  {\cal A}_{h,N}^{LL,LR}\, =\,
  m_{1}\,m_{2}\, {\cal I}_{h}\,
  {\phi}_{B_{c}^{\ast}}^{V}(x_{2})\,  \Big\{
  {\phi}_{D}^{a}(x_{3})\,
  (s+2\,m_{3}^{2}\,\bar{x}_{3})
  -{\phi}_{D}^{p}(x_{3})\,2\,m_{3}\,m_{b} \Big\}
   \label{fighN-LL},
   \end{equation}
  %------------------------------------
   \begin{equation}
  {\cal A}_{h,T}^{LL,LR}\, =\,
  2\, m_{1}\,m_{2}\, {\cal I}_{h}\,
  {\phi}_{B_{c}^{\ast}}^{V}(x_{2})\,
  {\phi}_{D}^{a}(x_{3})
   \label{fighT-LL},
   \end{equation}
  %------------------------------------
  %------------------------------------------------------------
  %------------------------------------
   \begin{equation}
  {\cal I}_{a}\ =\
  {\int}_{0}^{1}dx_{1}
  {\int}_{0}^{1}dx_{2}
  {\int}_{0}^{\infty}b_{1} db_{1}
  {\int}_{0}^{\infty}b_{2} db_{2}\,
   H_{a}({\alpha}_{e},{\beta}_{a},b_{1},b_{2})\,
   E_{a}(t_{a})\,
  {\alpha}_{s}(t_{a})
   \label{ca},
   \end{equation}
  %------------------------------------
  \begin{equation}
  {\cal I}_{b}\ =\
  {\int}_{0}^{1}dx_{1}
  {\int}_{0}^{1}dx_{2}
  {\int}_{0}^{\infty}b_{1} db_{1}
  {\int}_{0}^{\infty}b_{2} db_{2}\,
   H_{b}({\alpha}_{e},{\beta}_{b},b_{1},b_{2})\,
   E_{b}(t_{b})\,
  {\alpha}_{s}(t_{b})
   \label{cb},
   \end{equation}
  %------------------------------------
   \begin{eqnarray}
  {\cal I}_{c} &=&
   \frac{1}{N_{c}}
  {\int}_{0}^{1}dx_{1}
  {\int}_{0}^{1}dx_{2}
  {\int}_{0}^{1}dx_{3}
  {\int}_{0}^{\infty}db_{1}
  {\int}_{0}^{\infty}b_{2}db_{2}
  {\int}_{0}^{\infty}b_{3}db_{3}
   \nonumber \\ &{\times}&
   H_{c}({\alpha}_{e},{\beta}_{c},b_{2},b_{3})\,
   E_{c}(t_{c})\,
  {\alpha}_{s}(t_{c})\,
  {\delta}(b_{1}-b_{2})\,
   \label{cc},
   \end{eqnarray}
  %------------------------------------
   \begin{eqnarray}
  {\cal I}_{d} &=&
   \frac{1}{N_{c}}
  {\int}_{0}^{1}dx_{1}
  {\int}_{0}^{1}dx_{2}
  {\int}_{0}^{1}dx_{3}
  {\int}_{0}^{\infty}db_{1}
  {\int}_{0}^{\infty}b_{2}db_{2}
  {\int}_{0}^{\infty}b_{3}db_{3}
   \nonumber \\ &{\times}&
   H_{d}({\alpha}_{e},{\beta}_{d},b_{2},b_{3})\,
   E_{d}(t_{d})\,
  {\alpha}_{s}(t_{d})\,
  {\delta}(b_{1}-b_{2})\,
   \label{cd},
   \end{eqnarray}
  %------------------------------------
   \begin{eqnarray}
  {\cal I}_{e} &=&
   \frac{1}{N_{c}}
  {\int}_{0}^{1}dx_{1}
  {\int}_{0}^{1}dx_{2}
  {\int}_{0}^{1}dx_{3}
  {\int}_{0}^{\infty}b_{1}db_{1}
  {\int}_{0}^{\infty}b_{2}db_{2}
  {\int}_{0}^{\infty}db_{3}
   \nonumber \\ &{\times}&
  H_{e}({\alpha}_{a},{\beta}_{e},b_{1},b_{2})\,
  E_{e}(t_{e})\,
  {\alpha}_{s}(t_{e})\,
  {\delta}(b_{2}-b_{3})
   \label{ce},
   \end{eqnarray}
  %------------------------------------
   \begin{eqnarray}
  {\cal I}_{f} &=&
   \frac{1}{N_{c}}
  {\int}_{0}^{1}dx_{1}
  {\int}_{0}^{1}dx_{2}
  {\int}_{0}^{1}dx_{3}
  {\int}_{0}^{\infty}b_{1}db_{1}
  {\int}_{0}^{\infty}b_{2}db_{2}
  {\int}_{0}^{\infty}db_{3}
   \nonumber \\ &{\times}&
  H_{f}({\alpha}_{a},{\beta}_{f},b_{1},b_{2})\,
  E_{f}(t_{f})\,
  {\alpha}_{s}(t_{f})\,
  {\delta}(b_{2}-b_{3})
   \label{cf},
   \end{eqnarray}
  %------------------------------------
   \begin{equation}
  {\cal I}_{g}\ =\
  {\int}_{0}^{1}dx_{2}
  {\int}_{0}^{1}dx_{3}
  {\int}_{0}^{\infty}b_{2} db_{2}
  {\int}_{0}^{\infty}b_{3} db_{3}\,
   H_{g}({\alpha}_{a},{\beta}_{g},b_{2},b_{3})\,
   E_{g}(t_{g})\,
  {\alpha}_{s}(t_{g})
   \label{cg},
   \end{equation}
  %------------------------------------
   \begin{equation}
  {\cal I}_{h}\ =\
  {\int}_{0}^{1}dx_{2}
  {\int}_{0}^{1}dx_{3}
  {\int}_{0}^{\infty}b_{2} db_{2}
  {\int}_{0}^{\infty}b_{3} db_{3}\,
   H_{h}({\alpha}_{a},{\beta}_{h},b_{2},b_{3})\,
   E_{h}(t_{h})\,
  {\alpha}_{s}(t_{h})
   \label{ch},
   \end{equation}
  %------------------------------------
  %-----------------------------------------------------
  where $\bar{x}_{i}$ $=$ $1$ $-$ $x_{i}$ and $x_{i}$
  are longitudinal momentum fraction of valence quarks;
  $b_{i}$ is the conjugate variable of the
  transverse momentum $k_{iT}$;
  Sudakov factors $E_{i}$ are defined as
  %-----------------------------------------------------
   \begin{equation}
   E_{i}(t) =
   \left\{ \begin{array}{lll}
  {\exp}\{ -S_{{\Upsilon}}(t)-S_{B_{c}^{\ast}}(t) \}, &~& i=a,b \\
  {\exp}\{ -S_{{\Upsilon}}(t)-S_{B_{c}^{\ast}}(t)-S_{D}(t) \}, & & i=c,d,e,f \\
  {\exp}\{ -S_{B_{c}^{\ast}}(t)-S_{D}(t) \}, & & i=g,h
   \end{array} \right.
   \label{sudakov-exp},
   \end{equation}
  %-----------------------------------------------------
  %-----------------------------------------------------
   \begin{eqnarray}
   S_{{\Upsilon}}(t) &=&
   s(x_{1},p_{1}^{+},1/b_{1})
  +2{\int}_{1/b_{1}}^{t}\frac{d{\mu}}{\mu}{\gamma}_{q}
   \label{sudakov-bb}, \\
  %-----------------------------------------------------
   S_{B_{c}^{\ast}}(t) &=&
   s(x_{2},p_{2}^{+},1/b_{2})
  +2{\int}_{1/b_{2}}^{t}\frac{d{\mu}}{\mu}{\gamma}_{q}
   \label{sudakov-bc}, \\
  %-----------------------------------------------------
   S_{D}(t) &=&
   s(x_{3},p_{3}^{+},1/b_{3})
  +2{\int}_{1/b_{3}}^{t}\frac{d{\mu}}{\mu}{\gamma}_{q}
   \label{sudakov-ds}.
   \end{eqnarray}
  %-----------------------------------------------------
  The definition of functions $H_{i}$ and
  scale $t_{i}$ are the same as that of Ref.\cite{plb752}.
  \end{appendix}

   %%%%%%%%%%%%%%%%%%%%%%%%%%%%%%%%%%%%%%%%%%%%%%%%%%%%%%%%%%%
  
  \end{document}